\documentclass[twocolumn,amsmath,amssymb,aps,superscriptaddress,longbibliography,10pt]{revtex4-2}
\usepackage{graphicx}
\usepackage{color}
\usepackage{bm}
\usepackage{dcolumn}
\usepackage{multirow}
\usepackage{amsmath}
\usepackage{amssymb}
\usepackage{amscd}
\usepackage{theorem}
\usepackage[mathcal]{eucal}
\usepackage{mathptmx}
\usepackage[abs]{overpic}
\usepackage{caption}
\usepackage{subcaption}
\usepackage{chemformula}
\usepackage{comment}

\begin{document}

\title[]{\textit{Ab-initio} study of structural, vibrational and non-linear optical properties of (TiO$_2$)-(Tl$_2$O)-(TeO$_2$) glasses}
\author{Raghvender Raghvender}
\author{Assil Bouzid}
\email{assil.bouzid@cnrs.fr}
\affiliation{~Institut de Recherche sur les C\'eramiques (IRCER), UMR CNRS 7315-Universit\'e de Limoges, France.}
\author{Evgenii M. Roginskii}
\affiliation{~Ioffe Institute, Polytekhnicheskaya 26, 194021, St. Petersburg, Russia}
\author{David Hamani}
\affiliation{~Institut de Recherche sur les C\'eramiques (IRCER), UMR CNRS 7315-Universit\'e de Limoges, France.}
\author{Olivier Noguera}
\affiliation{~Institut de Recherche sur les C\'eramiques (IRCER), UMR CNRS 7315-Universit\'e de Limoges, France.}
\author{Philippe Thomas}
\affiliation{~Institut de Recherche sur les C\'eramiques (IRCER), UMR CNRS 7315-Universit\'e de Limoges, France.}
\author{Olivier Masson}
\email{olivier.masson@unilim.fr}
\affiliation{~Institut de Recherche sur les C\'eramiques (IRCER), UMR CNRS 7315-Universit\'e de Limoges, France.}

\date{\today}

\begin{abstract}

This paper reports on a  systematic first-principles molecular dynamics investigation of binary $\rm{(TlO_{0.5})}_{\textit{y}}-{(TeO_2)}_{1-\textit{y}}$ and ternary $\rm{(TiO_{2})}_{\textit{x}}-{(TlO_{0.5})}_{\textit{y}}-{(TeO_2)}_{1-\textit{x}-\textit{y}}$ tellurite glasses. The obtained structural models are validated against available measured X-ray pair distribution functions. In the binary system, increasing TlO$_{0.5}$ content induces network depolymerization through the reduction of Te coordination number, the substitution of Te–O–Te linkages with Te=O$^{-}\cdots$Tl$^{+}$ units, and the proliferation of non-bridging oxygens. In addition, rings analysis demonstrates a loss of the network connectivity via the opening of small n-membered rings. In contrast, TiO$_2$ acts as a network former in ternary glasses, preserving Te coordination number, and promoting a high fraction of bridging oxygens. Ti atoms induces a network repolymerization that manifests through the formation of smaller Ti-containing n-membered rings thereby balancing the strong effect of Tl$_2$O modifier. Beside the structural analysis, we also computed Raman spectra and non-linear optical properties on the obtained large periodic models. Our results reproduce experimental trends in Raman band shifts with composition, while nonlinear optical calculations show that $\langle\chi^{(3)}\rangle$ remains stable with TlO$_{0.5}$ addition in binary glasses, consistent with experiment. In the case of ternary systems, we find that the inclusion of a small fraction of TiO$_2$ preserves the high optical nonlinearity of the TeO$_2$ network while maintaining the overall network connectivity. These results establish a predictive framework for tailoring the atomic structure and nonlinear optical response of tellurite glasses through the controlled interplay of  modifiers nature and concentration.

\end{abstract}

\maketitle

\section{Introduction}
TeO$_2$-based materials (tellurite materials) exhibit outstanding non-linear optical (NLO) properties, making them highly suitable for applications such as optical switching devices, amplifiers, frequency up-conversion systems, and laser hosts \cite{rivera2017technological,el2018tellurite}.
In particular, TeO$_2$-based glasses possess a high third-order NLO susceptibility compared to other oxide glasses \cite{el2001tellurite}.
Nevertheless, pure TeO$_2$ forms a glass only under specific conditions \cite{conditional_glass} and requires rapid quenching rates to achieve stable amorphous samples \cite{brady,garaga,ELDEEN2008333,ELKHOSHKHANY2020119994,KAKY2017166, TAGIARA2017116}. Therefore, TeO$_2$ is often mixed with one or more modifier oxides (MO) \cite{el2001tellurite} to achieve stable glasses.

The atomic-scale structure of TeO$_2$ based glasses consists in a majority of TeO$_4$ disphenoids and TeO$_3$ trigonal pyramids that contain bonding oxygens (BO) and/or non-bonding oxygens (NBO, or terminal oxygens) \cite{pietrucci2008teo,barney2013terminal,gulenko_atomistic_2014,kalampounias2015glass,marple2019structure,raghvender2022structure,khanna2024structure}. Each structural unit possesses an electronic lone pair (LP) near the Te$^{4+}$ cation, and these units are interconnected by $-$Te$-$O$-$Te$-$ bridges. TeO$_4$ units mainly constitute the glassy network whereas TeO$_3$ units can be more isolated and thus contain a higher population of NBO. Introducing MO into the amorphous TeO$_2$ matrix breaks the $-$Te$-$O$-$Te$-$ linkages and transforms TeO$_4$ into TeO$_3$ units, which reduces the coordination number of Te$^{4+}$ and promotes structural depolymerization \cite{el-mallawany_tellurite_2002,christy2016review, zemann1968crystal, zemann1971contributions,tagg1994crystal, hussin2009structural}.
This typically results in a decline in the NLO properties of the material \cite{mirgorodsky2006ab,hamami,manning2012ternary,Roginskii2017,roginskii2024nonlinear,roginskii2023ab}. Among oxide materials, thallium oxide (Tl$_2$O) is particularly of great interest as a modifier oxide, as it preserves the high NLO properties of TeO$_2$-based glasses \cite{jeansannetas_glass_1999, dutreilh2003new} even though it promotes structural depolymerization, as indicated by Raman spectroscopy studies \cite{sekiya, udovic2009formation, NOGUERA2004981}. Thallium, the rarest element in the boron group, exists in two oxide forms: thallous oxide (Tl$_2$O) and thallic oxide (Tl$_2$O$_3$). Although, we are not aware of any studies that report the use of Tl$_2$O$_3$ as an effective modifier oxide, Tl$_2$O has been widely used and studied to tailor the properties of TeO$_2$-based glasses \cite{jeansannetas_glass_1999, NOGUERA2004981, SOULIS2008143, dutreilh2003new}.

While the addition of Tl$_2$O improves the glass-forming ability of TeO$_2$-based glasses, the resulting binary glass exhibits limited thermal stability \cite{noguera2004dynamics,udovic2009formation}. Studies indicate that incorporating TiO$_2$ into the TeO$_2$ matrix enhances the thermal tolerance of the glass and significantly boosts its mechanical strength \cite{sabadel1997structural, udovic2006thermal, sabadel1999mossbauer, torzu2020}. Further, Ti$^{4+}$ and Te$^{4+}$ feature the same charge state and similar bond lengths to O, indicating that neither element acts strictly as a modifier relative to the other \cite{SOULIS2008143} This interpretation is supported by Dietzel's field strength criteria \cite{Dietzel19429,dietzel2013emaillierung}, which classify Ti$^{4+}$ as an intermediate glass former, with a field strength similar to that of Te$^{4+}$. Additionally, Ti$^{4+}$ is proposed to inhibit the formation of TeO$_3$ structural units in the TiO$_2$-TeO$_2$ glassy system, resulting in a well-polymerized glass network \cite{sabadel1997structural}. The enhanced mechanical resistance of the glass network can be attributed to the presence of rigid TiO$_6$ octahedra, which forms $-$Te$-$O$-$Ti$-$ linkages within the structure \cite{sabadel1997structural, sabadel1999mossbauer, udovic2006thermal}. Consequently, adding a small concentration of TiO$_2$ to thallium tellurite glasses has been recommended to achieve an optimal balance between mechanical and optical properties \cite{udovic2009formation}.

The local environment of Tl$^+$ in the amorphous matrix of TeO$_2$-based glasses remains largely unexplored. Specifically, there is a limited understanding of how Tl$^{+}$ cations bond with BOs and NBOs. While classical molecular dynamics simulations have been applied to study thallium tellurite glasses \cite{raghvender2022buckingham}, these methods lack an accurate electronic description due to their empirical nature. Experimental studies on Tl$_2$O-TeO$_2$  and TiO$_2$-Tl$_2$O-TeO$_2$ glasses have thus primarily focused on macroscopic properties, with only Raman spectroscopy providing some insights into the atomic-scale organisation of the material \cite{NOGUERA2004981,udovic2009formation}.
At this stage, $\textit{ab-initio}$ modeling offers a promising approach in understanding the role of thallium and titanium oxides in shaping the structural and optical properties of Tl$_2$O-TeO$_2$ and TiO$_2$-Tl$_2$O-TeO$_2$ glasses. This is the main goal of the present work.

This paper is organized as follows: section II describes the computational methods and validation of the generated glassy models. Section III presents a detailed structural analysis, examining the atomic pair distribution functions (PDF), coordination numbers, local atomic environments, and ring statistics on Tl$_2$O-TeO$_2$ binary and TiO$_2$-Tl$_2$O-TeO$_2$ ternary glasses. In section IV, we discuss the computed Raman spectra and NLO properties of these glasses. Finally, conclusions are drawn in section V.

\section{Computational method and models validation}
\subsection{Computational method}
In what follows we refer to Tl$_2$O as TlO$_{0.5}$ when expressing the glass compositions. Using the PACKMOL code \cite{Packmol}, we generated random initial configurations of the binary $\mathrm{(TlO_{0.5})}_{y}-\mathrm{(TeO_2)}_{1-x-y}$ and ternary $\mathrm{(TiO_2)}_{x}-\mathrm{(TlO_{0.5})}_{y}-\mathrm{(TeO_2)}_{1-x-y}$ glassy systems by distributing randomly oriented $\mathrm{TiO_2}$, $\mathrm{Tl_2O}$, and $\mathrm{TeO_2}$ molecules within a periodic cubic simulation cell. As detailed in Table ~\ref{tab4_amorphous_description}, 11 distinct models were built with varying molar concentrations of $\mathrm{TlO_{0.5}}$ and $\mathrm{TiO_2}$ to investigate the effects of TlO$_{0.5}$ and TiO$_2$ incorporation on the structure of the glass compared to the parent $\mathrm{TeO_2}$ system.

\begin{table}[!htbp]
    \centering
    \caption{Compositions of $\mathrm{(TiO_2)}_{x}-\mathrm{(TlO_{0.5})}_{y}-\mathrm{(TeO_2)}_{1-x-y} $ amorphous systems with various molar concentrations in \textit{x} and \textit{y} and the corresponding number of atoms in each model. The experimental specific mass density $\rho$\cite{udovic2009formation} and the corresponding cubic simulation cell side are also provided.}    
    \begin{footnotesize}
        \begin{tabular}{c|c|c|c}
        \hline \hline
        Concentration ($x$ and $y$)         & Number of             &Density    & Simulation                             \\
                                            & of Ti/Tl/Te/O (total) & $\rho$($ \rm g.cm^{-3}$)         & cell side ({\AA}) \\
        \hline \hline
        $x$ = 0\%, $y$ = 10\%     & 0/16/144/296  (456)   & 5.85 \cite{torzu2020}                          & 19.57            \\
        $x$ = 0\%, $y$ = 20\%     & 0/32/128/272  (432)   & 6.12 \cite{torzu2020}                          & 19.47            \\
        $x$ = 0\%, $y$ = 30\%     & 0/48/112/248  (408)   & 6.40 \cite{torzu2020}                          & 19.38            \\
        $x$ = 0\%, $y$ = 40\%     & 0/72/108/252  (432)   & 6.68 \cite{torzu2020}                          & 20.07            \\
        $x$ = 0\%, $y$ = 50\%     & 0/90/90/225   (405)   & 6.95 \cite{torzu2020}                          & 20.00            \\
                                            &                                                        &                     &                  \\
        $x$ = 5\%, $y$ = 20\%     & 10/40/150/340 (540)   & 5.94 \cite{udovic2009formation}                & 21.03            \\
        $x$ = 5\%, $y$ = 30\%     & 10/60/130/310 (510)   & 6.21 \cite{udovic2009formation}                & 20.93            \\
        $x$ = 5\%, $y$ = 40\%     & 10/80/110/280 (480)   & 6.51 \cite{udovic2009formation}                & 20.82            \\
                                            &                                                        &                     &                  \\
        $x$ = 10\%, $y$ = 10\%   & 16/16/128/296 (456)   & 5.60 \cite{udovic2009formation}                & 19.52            \\
        $x$ = 10\%, $y$ = 20\%   & 20/40/140/340 (540)   & 5.92 \cite{udovic2009formation}                & 20.87            \\
        $x$ = 10\%, $y$ = 30\%   & 20/60/120/310 (510)   & 6.20 \cite{udovic2009formation}                & 20.78            \\

        \hline \hline
    \end{tabular}
    \end{footnotesize}
    \label{tab4_amorphous_description}
\end{table}

The electronic structure is solved through density functional theory as implemented in the CP2K software package \cite{vandevondele2005quickstep}. We employed the Gaussian and Plane Waves (GPW) approach to solve the Kohn-Sham equations. In particular, triple-zeta valence polarization (TZVP) Gaussian-type basis set \cite{vandevondele2007gaussian} are used for all atomic species to expand the Kohn-Sham orbitals, and an auxiliary plane wave basis set was defined with an energy cut-off of 450~Ry and a relative cut-off of 50~Ry. Core-valence interactions are described by norm-conserving pseudopotentials of the Goedecker-Teter-Hutter (GTH) type \cite{goedecker1996separable, hartwigsen1998relativistic} 
and the generalized gradient approximation (GGA) for the exchange-correlation (XC) functional was adopted within the PBE parameterization \cite{PBE}. 

First-principles molecular dynamics (FPMD) are performed within the Born-Oppenheimer molecular dynamics (BOMD) framework. The equations of motion were integrated with a time step of 1~fs and periodic boundary conditions were applied in the three directions. The simulations were conducted in the canonical (NVT) ensemble, maintaining a constant number of particles, simulation cell volume, and system temperature. Temperature control was achieved using a Nosé-Hoover thermostat \cite{NOSE, Nose1, Hoover}.

Glassy systems were produced through melt-quench protocol where all the systems were maintained at a temperature higher than the melting point until the diffusion coefficients of all atomic species demonstrated liquid-like diffusive behavior. Subsequently, the glassy state was achieved by quenching/cooling the system down to room temperature.
The applied thermal annealing cycle is as follows: 8 ps at T = 2000K, 9 ps at T = 1500 K, 25 ps at T = 1000 K, 25 ps at T = 600 K and 30 ps at T = 300 K. Statistacal averages of the structural properties were obtained by analysing the last 20~ps of the trajectory at room temperature.

In order to establish an electronic-structure-based investigation of the key structural parameters of the glassy network (coordination numbers and atomic local environments), we resort to the the maximally localized Wannier functions (MLWFs) formalism \cite{resta1999electron, marzari2012maximally}. 
In this scheme, a Wannier center gives access to the most probable spatial localization of shared electrons (constituting a bond) or lone pair electrons. Wannier orbitals ($w_{n}$(r)) and their corresponding centers are obtained by an on the fly unitary transformation of the Kohn–Sham orbitals $\psi_{i}$(r) under the constraint of minimizing the spatial extension (spread, $\Omega$) of the resulting $w_{n}$(r) as follows: 

\begin{eqnarray}
 \Omega = \sum_{n} \Big(\big< w_{n}|r^{2}|w_{n}\big>\Big)- \Big(\big< w_{n}|r|w_{n}\big>^{2}\Big). \label{eq0}
\end{eqnarray}

The resulting centers represent the localization of two electrons and indicates their average position. In particular, \ch{W^{B}} and \ch{W_{$\alpha$}^{LP}} stand for centers corresponding to chemical bonds and lone-pair electrons associated to ion $\alpha$, respectively. Within this formalism, if two atoms $\alpha$ and $\beta$ situated at a distance $d_{\alpha\beta}$ and sharing a Wannier center, located respectively at distances $d_{\alpha W}$ and $d_{\beta W}$, and that satisfy the inequality $|d_{\alpha\beta}-d_{\alpha W}-d_{\beta W}| \le 0.05$ {\AA}, they are considered to be bonded \cite{raghvender2022structure,shuaib2025atomic}. A tolerance of 0.05 {\AA} is taken into consideration to account for the deviations in the spatial localization of the center. Additionally, it has been shown that it is necessary to define a cutoff angle between \ch{W_{Te}^{LP}}, \ch{Te}, and \ch{O} to be $\ge$ 73$^\circ$ in order to dismiss lone pair Wannier centers that may arise at distances that fulfill the bonds inequality requirement but do not correspond to chemical bonds \cite{raghvender2022structure,shuaib2025atomic,gulenko2014atomistic}. 

Following this procedure, oxide anions are labeled as bridging (BO, e.g. forming Te-O-Te bridges) and non bridging (NBO, e.g. forming Te=O terminal bonds) based on their bonding nature with tellurium cations. Consequently, one can achieve a counting of the number of Te-O bonds, thereby reducing the effects of selecting a fixed bond-length cutoff and enabling a more accurate estimation of the coordination number of as well as a decomposition of the local environment around Te$^{4+}$ ions. In this work, Wannier functions are computed at the PBE0 hybrid functional level of theory \cite{perdew1996rationale,adamo1999toward} on top of 100 configurations selected along the last 20 ps of the trajectory at T = 300 K.

\subsection{Models validation}

\begin{figure}[!tbp]
    \includegraphics[width=0.99\linewidth,keepaspectratio=true]{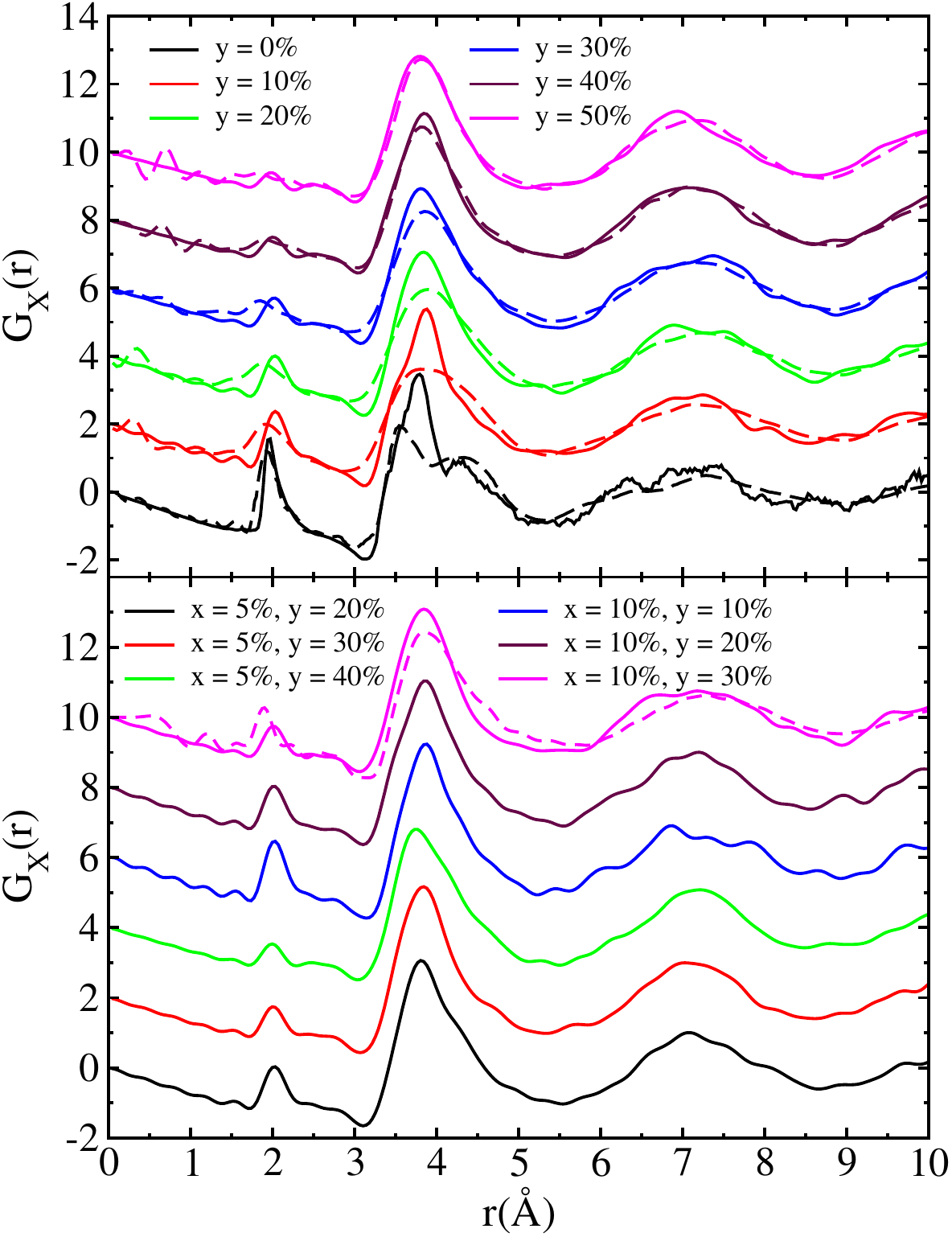}
    \caption{Comparison of total X-ray scattering PDF G($r$) between experiments (dashed lines) \cite{torzu2020} and FPMD simulations (solid lines) for $\mathrm{(TlO_{0.5})}_{y}-\mathrm{(TeO_2)}_{1-y}$ binary and $\mathrm{(TiO_2)}_{x}-\mathrm{(TlO_{0.5})}_{y}-\mathrm{(TeO_2)}_{1-x-y}$ ternary glasses. A vertical shift is applied for clarity.}
    \label{label_figure1}
\end{figure}

We analyse the structures of our models and confront them to experimental results by comparing the calculated and measured X-ray total PDF $G_{X}(r)$, as shown in Figure~\ref{label_figure1}. $G_{X}(r)$ is usually defined and calculated as follows:

\begin{eqnarray}
G_{X}(r) \sim \sum_{\alpha=1}^{3}\sum_{\beta = 1}^{3}\frac{c_{\alpha}c_{\beta}f_{\alpha}(q_{0})f_{\beta}(q_{0})}{|<f(q_{0})>^{2}|}\Big[g_{\alpha\beta}(r)-1\Big].
\end{eqnarray}

where, $g_{\alpha\beta}(r)$ is the partial pair distribution function for species $\alpha$ and $\beta$ and $f$ their respective scattering factors evaluated at an arbitrary $q_0$ value typically set to $q_0 = 0$. In this work, we instead employed the exact expression for ${\rm G}(r)$ from Ref. (\citenum{masson2013exact}), which involves a weighted linear combination of modified partial pair correlation functions. The determination of the weights is based on the mean values of the Faber-Ziman factors over the range of the considered reciprocal space. Subsequently, the PDF is transformed to reciprocal space and back transformed to real space using the experimental Q$_{max}$ value of 17 {\AA}$^{-1}$. This procedure, ensures a fair comparison of the calculated PDFs to the experimental ones by guaranteeing similar data treatment.

Compared to the four available experimental PDFs for $\mathrm{(TlO_{0.5})}_{y}-\mathrm{(TeO_2)}_{1-y}$ binary glasses and to the one available for the $\mathrm{(TiO_2)}_{x}-\mathrm{(TlO_{0.5})}_{y}-\mathrm{(TeO_2)}_{1-x-y}$ ternary glassy system, the simulated PDFs show an overall good agreement.
We observe that the peak corresponding to Te–O bond distances around 2.0 {\AA}, is slightly shifted to higher distances by approximately 0.1 {\AA} compared to the experimental data. This shift can be attributed to the GGA functional and GTH pseudopotentials, which are known to overestimate bond lengths \cite{shuaib2025atomic}.
In addition, this agreement is less good for models at low \ch{TlO_{0.5}} concentrations, particularly when looking at the intensity of the main peak located around 4 {\AA}. It has been shown in a previous work on pure $\mathrm{TeO_2}$, that the PBE0 hybrid functional led to a more accurate reproduction of this particular peak but with a minimal effect on the structure of the glass \cite{raghvender2022structure}. 
In fact, hybrid functional describes more accurately electron localization of the lone pair on the Te$^{4+}$ cation, thereby ensuring a better estimation of Te--Te correlations than PBE functional. Due to the prohibitive high computational cost of hybrid functional molecular dynamics, and the fact that necessity for hybrid functionals decreases with lesser tellurium content in the system, we stick to the PBE-GGA functional based molecular dynamics in the present study to simulate binary and ternary glasses. 
Overall, in both binary and ternary systems, all the major experimental PDF features and their evolution as a function of the composition are reproduced, thereby indicating that our models successfully capture the structural reorganization in the glass across varying compositions. 

\section{Results and discussion}
\subsection{Structural analysis of the binary glasses} \label{Structural analysis}
\subsubsection{Pair distribution functions}

\begin{figure}[!htbp]
    \centering
    \includegraphics[width=0.99\linewidth,keepaspectratio=true]{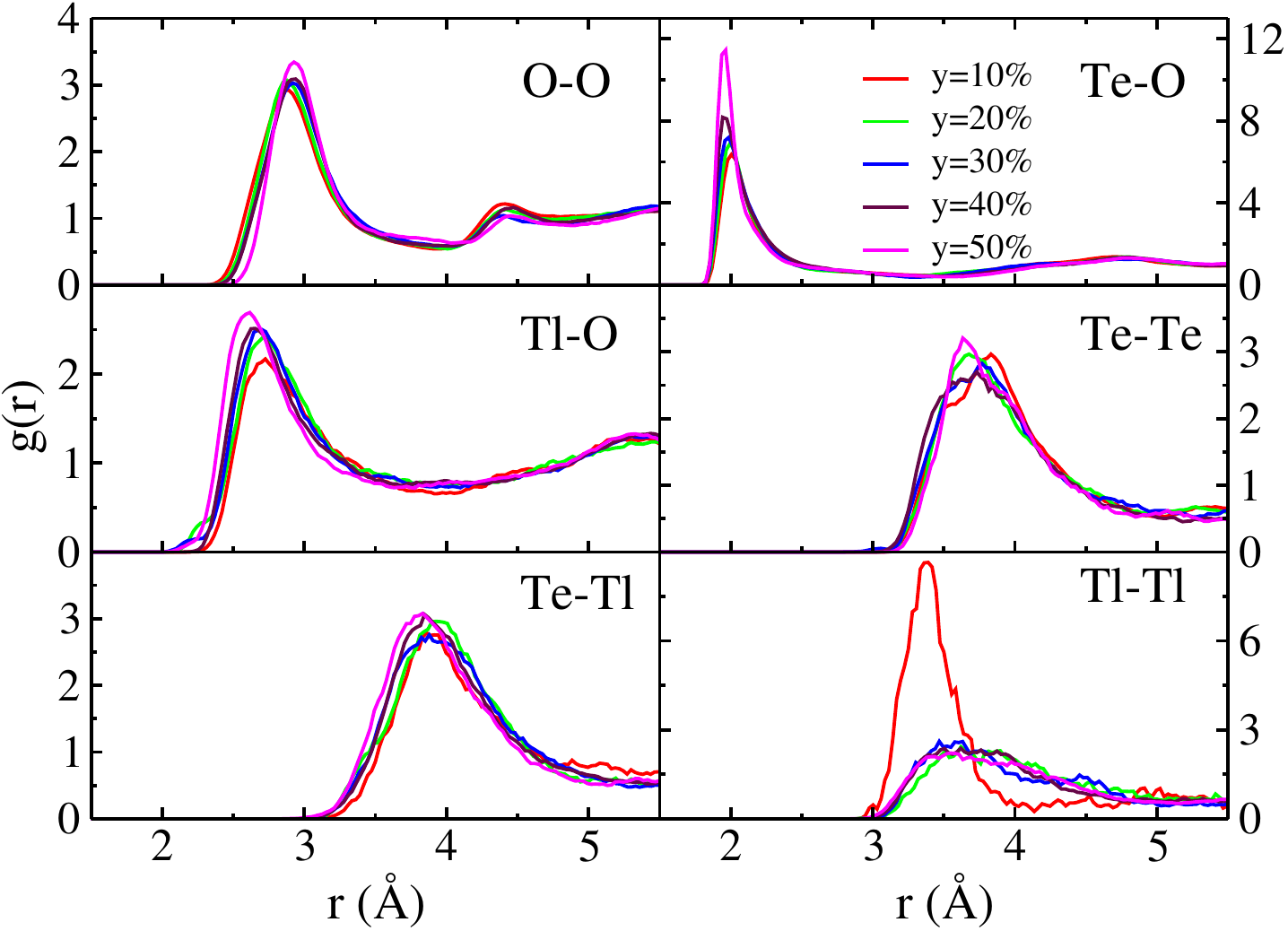}
    \caption{Partial PDFs g$_\text{O-O}$($r$), g$_\text{Te-O}$($r$), g$_\text{Tl-O}$($r$), g$_\text{Te-Te}$($r$), g$_\text{Te-Tl}$($r$) and g$_\text{Tl-Tl}$($r$) in binary $\mathrm{(TlO_{0.5})}_{y}-\mathrm{(TeO_2)}_{1-y}$ glasses with various $y$ concentrations.}
    \label{label_figure2}
\end{figure}

\begin{figure}[!htbp]
    \centering
    \includegraphics[width=0.9\linewidth,keepaspectratio=true]{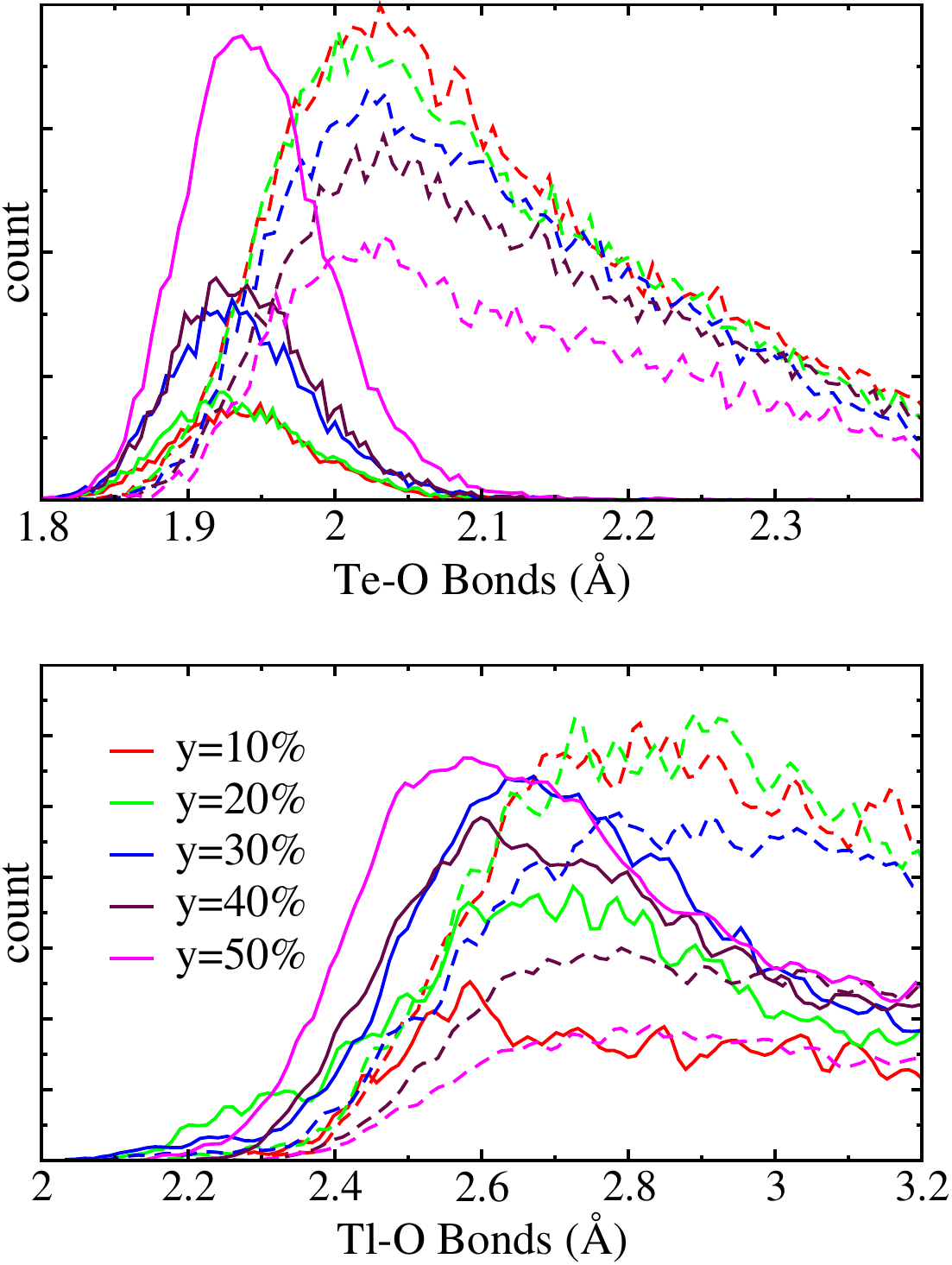}
    \caption{Distribution of bond-distances in Te (Top panel) and Tl (bottom panel) chemical groups with bridging oxygen (dashed lines) and non-bridging oxygen (solid lines). The distribution was normalized to the number of cations in each system.}
    \label{label_figure3}
\end{figure}

The short-range correlation in binary systems was first analysed by calculating the partial PDFs g$_{\alpha \beta}(r)$ as illustrated in Figure \ref{label_figure2}. First, a slight shift of the first peak in g$_\text{Te-O}$($r$) and g$_\text{Tl-O}$($r$) toward smaller $r$ values is observed when increasing $y$. Te$-$NBO pairs generally have shorter average bond distances than Te$-$BO pairs to balance the bond valence of the Te$^{4+}$ cation. The Te$-$BO and Te$-$NBO distance distributions plotted in Figure \ref{label_figure3} show that the Te$-$NBO contribution increases with higher $\mathrm{TlO_{0.5}}$ concentrations and exhibit average bond distances approximately 0.1 {\AA} shorter than those of Te$-$BO pairs. In addition, neither Te$-$BO nor Te$-$NBO peaks show a significant shift in their individual positions across different concentrations. Therefore, the overall shift to shorter distances of the Te$-$O peak in the partial PDF (see Figure~\ref{label_figure2}) is primarily due to the increasing population of Te$-$NBO pairs. Similar trends are observed in the case of the Tl$-$O partial PDF (see Figure~\ref{label_figure2}) with the first peak position shifting from 2.74 {\AA} at $y = 10\%$ to 2.61 {\AA} at $y = 50\%$. These results align also with the shift observed in the first peak of the experimental G(r) (see Figure~\ref{label_figure1}). 

Second, as the concentration of the $\mathrm{TlO_{0.5}}$ modifier increases, the intensity of the Te$-$BO and Tl$-$BO bond distributions decrease while the intensity of the Te$-$NBO and Tl$-$NBO bond distributions increase (see Figure \ref{label_figure3}). The latter have a more symmetric and narrower distribution, resulting in an overall increase in the intensities of partial g$_\text{Te-O}$($r$) and g$_\text{Tl-O}$($r$) PDFs (see Figure \ref{label_figure2}). 

As for the g$_\text{O-O}(r)$ partial PDF (see Figure~\ref{label_figure2}), the addition of thallium oxide to the \ch{TeO2} glass network leads to a slight shift of the first peak, moving from approximately 2.88 {\AA} at $y = 10\%$ to 2.93 {\AA} at $y = 50\%$. The contribution of BO$-$BO pairs decreases with the addition of $\mathrm{TlO_{0.5}}$ (see Figure S1), while those of the NBO$-$NBO and NBO$-$BO pairs increase. Additionally, NBO$-$NBO pairs generally exhibit their peak at larger distances than BO$-$BO pairs. This observation aligns with previous findings \cite{raghvender2022structure} which showed that NBOs carry higher negative charges than BOs, leading to greater self-repulsion and thus, larger NBO$-$NBO distances. Consequently, the rise in NBO$-$NBO populations and their tendency to appear at larger distances than BO$-$BO pairs explains the observed shift in the O$-$O partial PDF (see Figure~\ref{label_figure2}). 

Regarding the cation$-$cation correlations, Figure~\ref{label_figure2} shows that the addition of $\mathrm{TlO_{0.5}}$ has a minimal impact on the peak positions in the partial PDFs for Te$-$Te and Te$-$Tl pairs. The first peak of the Te$-$Te partial PDF remains consistently positioned around 3.7 {\AA} across all compositions, closely matching the peak position observed in pure TeO$_2$ glass \cite{raghvender2022structure}. For the Tl$-$Tl correlation, a similar trend is seen across different compositions, with the exception of $y = 10\%$ where the limited number of Tl$^{+}$ cations in the model introduces large statistical fluctuations. 

\subsubsection{Average coordination numbers}

We now assess the coordination numbers of Te$^{4+}$ and Tl$^{+}$ cations.
The coordination number of tellurium, $n_{\text{Te}}$, cannot be precisely determined due to the ill description of the first minimum in the Te$-$O partial PDF \cite{raghvender2022structure}. Therefore, we resort to the MLWFs formalism to better describe the complex bonding environment around Te atoms. Figure S2 shows the distribution of Te$-$W and O$-$W distances for $\mathrm{(TlO_{0.5})}_{y}-\mathrm{(TeO_2)}_{1-y}$ glasses. The Wannier centers (W) associated with the Te and O lone-pair electrons are represented by the first peak centered at approximately 0.5 {\AA} for Te$-$W and 0.32 {\AA} for O$-$W pair distance distributions. The second peak located in the range 1.2-1.81 {\AA} for Te$-$W and 0.4-0.65 {\AA} for O$-$W pair distance distributions, correspond to bonding W centers occurring along the Te-O bonds. By summing up the second minimum position in both distributions, we determine a suitable cut-off distance (r$_{\text{cut-off}}$) for characterizing the Te$-$O bond equal to be 2.46 {\AA} across all compositions of binary glasses. Identical results were obtained for pure TeO$_2$ \cite{raghvender2022structure}. 
In the case of Tl-O linkages, the bonding Wannier centers are localized on the oxide anions, making it challenging to distinguish these centers from those corresponding to the lone pairs on oxide anions. Consequently, the coordination number of thallium, n$_{\text{Tl-O}}$, is then obtained by integrating the partial PDF g$_{\text{Tl-O}}(r)$. However as the Tl-O partial PDF does not feature a clear minimum, we estimate n$_{\text{Tl-O}}$ using the longest Tl$-$O bond found in Tl-Te-O crystalline structures. Specifically a value of 3.26 {\AA} found in $\alpha$-Tl$_2$Te$_2$O$_5$\cite{jeansannetas1998crystal} is used.

\begin{figure}[!htbp]
    \centering
    \includegraphics[width=0.99\linewidth,keepaspectratio=true]{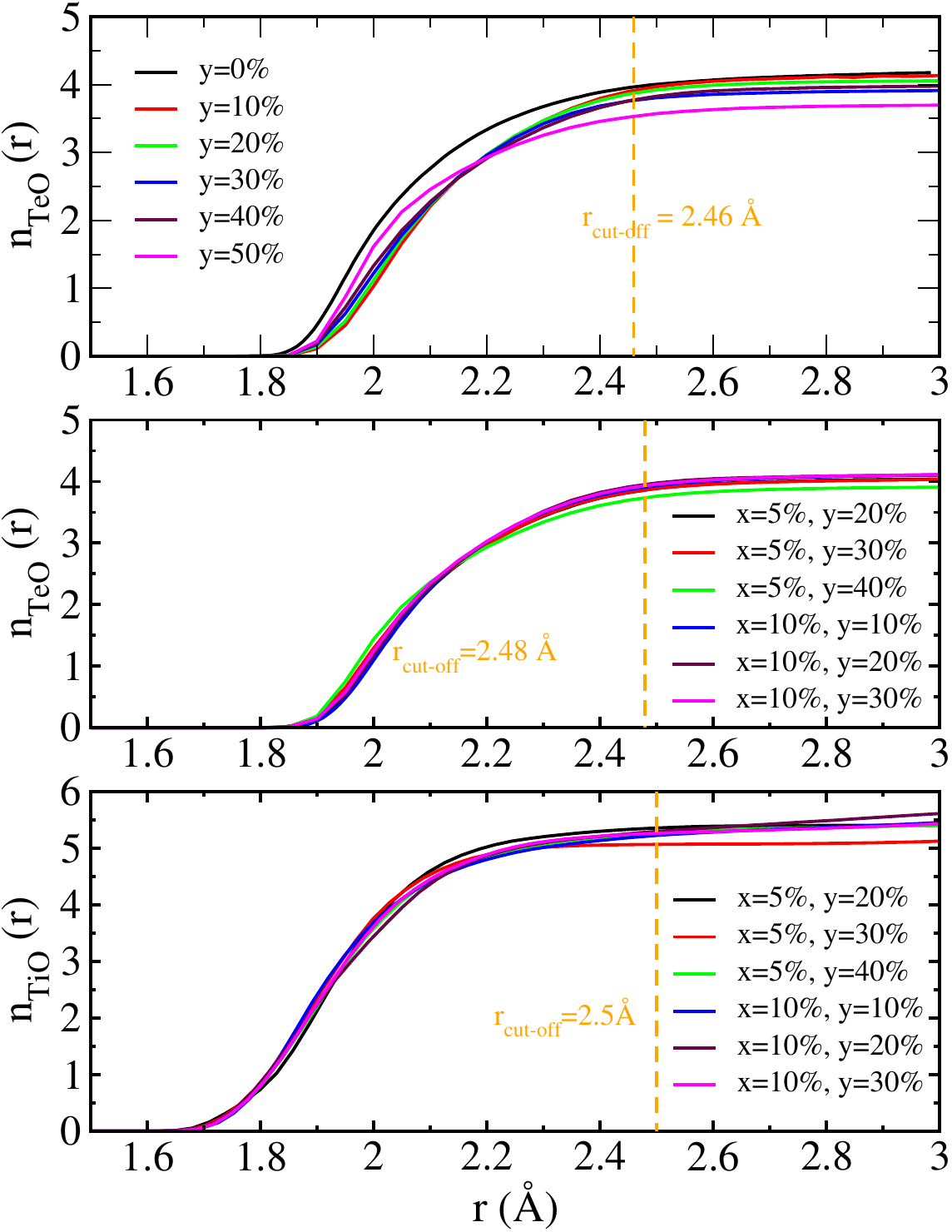}
        \caption{(Top panel) Running coordination number $n_{\text{TeO}}$(r) as a function of Te$-$O pair distance for all considered $\mathrm{(TlO_{0.5})}_{y}-\mathrm{(TeO_2)}_{1-y}$ binary glasses, computed using the MLWF formalism. 
        Running coordination numbers (middle panel) $n_{\text{TeO}}$(r) computed using the MLWF formalism, and (bottom panel) $n_{\text{TiO}}$(r) computed as the integral of the corresponding partial PDF, for $\rm  {(TiO_{2})}_{\textit{x}}-{(TlO_{0.5})}_{\textit{y}}-{(TeO_2)}_{1-\textit{x}-\textit{y}}$ ternary glasses.
        Vertical dashed lines indicate distance cut-offs used to compute the average coordination numbers.}
    \label{label_figure4}
\end{figure}

The running coordination number n$_{\text{Te-O}}$(r) is plotted as a function of the radial distance $r$ in Figure \ref{label_figure4}. The coordination curve increases with increasing $r$ until it reaches a plateau at around the established cut-off distance $r_{\text{cut-off}}$=2.46 {\AA}. The average coordination number $n_{\text{Te}}$ obtained using this value are reported in Table ~\ref{tab4_coordination_various_glass}. Incorporating $\mathrm{TlO_{0.5}}$ in the glass network leads to a reduction in $n_{\text{Te}}$ from 3.91 to 3.53 when going from $y = 10\%$ to $y = 50\%$. 

\begin{table}[!htbp]
    \centering
\caption{Coordination number n$_{\text{Te-O}}$ in $\rm {(TlO_{0.5})}_{\textit{y}}-{(TeO_2)}_{1-\textit{y}}$ binary glasses and Tl-Te-O crystals. }    
    \begin{tabular}{r|l}
        \hline \hline
        Concentration ($y$)               & n$_{\text{Te-O}}$  \\
        \hline \hline
        $y$ = 10\%                        & 3.91 $\pm$ 0.04                                      \\
        $y$ = 20\%                        & 3.88 $\pm$ 0.04                                      \\
        $y$ = 30\%                        & 3.77 $\pm$ 0.04                                      \\
        $y$ = 40\%                        & 3.77 $\pm$ 0.04                                      \\
        $y$ = 50\%                        & 3.53 $\pm$ 0.04                                      \\
        \hline
        Crystals                          & n$_{\text{Te-O}}$   \\
        \hline
        Tl$_2$Te$_3$O$_7$ ($y$ = 40\%)          & 3.72 $\pm$ 0.06                                      \\
        $\alpha$-Tl$_2$Te$_2$O$_5$ ($y$ = 50\%) & 3.76 $\pm$ 0.11                                      \\
        Tl$_2$TeO$_3$  ($y$ = 67\%)             & 3.01 $\pm$ 0.02                                      \\
        \hline \hline
    \end{tabular}
    \label{tab4_coordination_various_glass}
\end{table}

\begin{figure}[!htbp]
    \centering
    \includegraphics[width=0.99\linewidth,keepaspectratio=true]{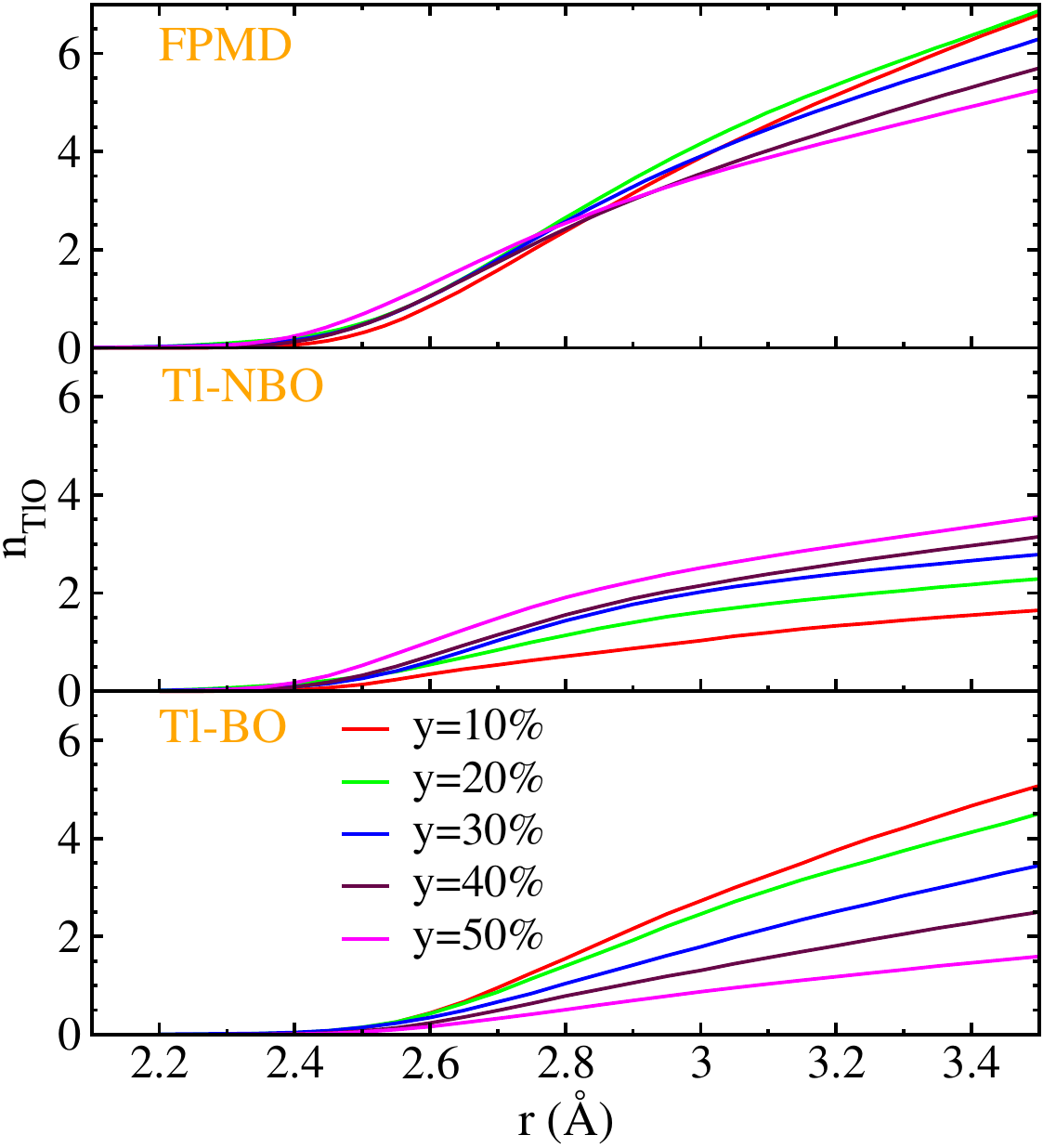}
    \caption{Tl running coordination numbers for the binary $\mathrm{(TlO_{0.5})}_{y}-\mathrm{(TeO_2)}_{1-y}$ glasses. The top panel shows the coordination number $n_{\text{Tl}}$ as a function of distance, obtained via integration of the partial Tl–O PDF. The middle panel displays the running $n_{\text{Tl-NBO}}$ for non-bridging oxygen (NBO), while the bottom panel illustrates the running $n_{\text{Tl-BO}}$ for bridging oxygen (BO).}
    \label{label_figure5}
\end{figure}

The coordination number of thallium, n$_{\text{Tl-O}}$, obtained by integrating the partial PDF g$_{\text{Tl-O}}(r)$ up to 3.26 {\AA}, as well as the decomposition of n$_{\text{Tl-O}}$ into contributions from n$_{\text{Tl-BO}}$ and n$_{\text{Tl-NBO}}$
is shown in the of Figure \ref{label_figure5}. Note that the terms BO and NBO refer to Te$^{4+}$ environments. The corresponding coordination numbers are provided in Table~\ref{tab4_Tl-O_Tl-NBO_coordination}.
Overall, we observe a reduction of n$_{\text{Tl-O}}$ from 5.44 at $y=10\%$ to 4.40 at $y=50\%$. In addition, the decomposition of the Tl coordination with respect to BO and NBO, show that this reduction is the result of a strong decrease in n$_{\text{Tl-BO}}$ and an increase of n$_{\text{Tl-NBO}}$.
This behaviour suggests that Tl$^{+}$ cations tend to be in a spatial proximity of NBO sites driven by the charge compensation of the negative charge on NBO site \cite{raghvender2022structure}, as illustrated in Figure~\ref{label_figure3}.

\begin{figure}[!htbp]
    \centering
    \includegraphics[width=0.99\linewidth,keepaspectratio=true]{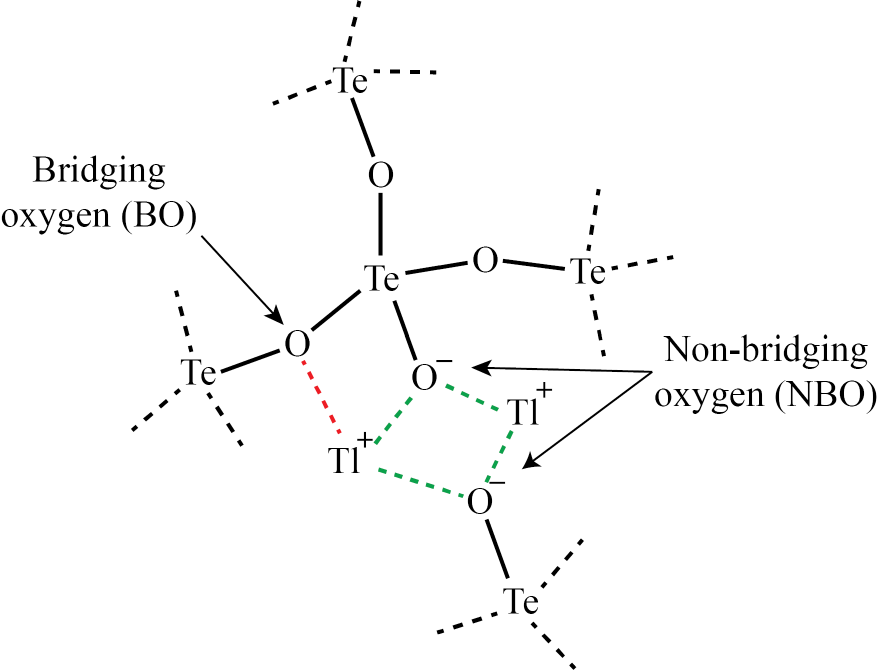}
    \caption{Schematic illustration of possible Tl$^+$ cation linkages with bridging oxygen (red color) and non-bridging oxygen (green lines).}
    \label{label_figure6}
\end{figure}

In silicate-based glasses \cite{NESBITT2015139, DU200466, CORMACK2003147, mead2006molecular, tilocca2006structural, kargl2006formation, machacek2010md}, three types of oxygen linkages are commonly identified: Si$-$BO$-$Si, Si$-$NBO$-$M (where M is a modifier), and a third type known as M$-$BO linkages. The M$-$BO linkages involve an oxide anion linked with two silicon cations and a modifier cation, resulting in a triply connected bridging oxygen (BO). 
Interestingly, our analysis shows the existence of such Tl$-$BO connections in the binary $\mathrm{(TlO_{0.5})}_{y}-\mathrm{(TeO_2)}_{1-y}$ glasses, as illustrated in Figure \ref{label_figure6}. 

\begin{table}[!htbp]
    \centering
    \caption{Tl coordination number n$_{\text{Tl-O}}$ = n$_\text{Tl-BO}$ + n$_\text{Tl-NBO}$ for $\rm {(TlO_{0.5})}_{\textit{y}}-{(TeO_2)}_{1-\textit{y}}$ glasses and Tl-Te-O crystalss. }
    \begin{tabular}{r | c | c | c}
        \hline \hline
        Concentration ($y$)                & n$_{\text{Tl-O}}$ & n$_\text{Tl-BO}$ & n$_\text{Tl-NBO}$ \\
        \hline \hline
        $y$ = 10\%                         & 5.44 $\pm$ 0.20     & 4.04 $\pm$ 0.23      & 1.40 $\pm$ 0.17       \\
        $y$ = 20\%                         & 5.59 $\pm$ 0.16     & 3.59 $\pm$ 0.16      & 2.00 $\pm$ 0.14       \\
        $y$ = 30\%                         & 5.18 $\pm$ 0.13     & 2.70 $\pm$ 0.14      & 2.48 $\pm$ 0.10       \\
        $y$ = 40\%                         & 4.67 $\pm$ 0.08     & 1.95 $\pm$ 0.09      & 2.72 $\pm$ 0.08       \\
        $y$ = 50\%                         & 4.40 $\pm$ 0.07     & 1.27 $\pm$ 0.07      & 3.13 $\pm$ 0.07       \\
        \hline
        Crystals                           &                 &                  &                   \\
        \hline
        Tl$_2$Te$_3$O$_7$ ($y$ = 40\%)           & 5.03 $\pm$ 0.11     & 1.46 $\pm$ 0.11      & 3.58 $\pm$ 0.08       \\
        $\alpha$-Tl$_2$Te$_2$O$_5$  ($y$ = 50\%) & 5.36 $\pm$ 0.11     & 2.25 $\pm$ 0.22      & 3.10 $\pm$ 0.24       \\
        Tl$_2$TeO$_3$ ($y$ = 67\%)               & 4.67 $\pm$ 0.08     & 0.04 $\pm$ 0.01      & 4.64 $\pm$ 0.09       \\
        \hline \hline
    \end{tabular}
    \label{tab4_Tl-O_Tl-NBO_coordination}
\end{table}

\begin{figure}[!htbp]
    \centering
    \includegraphics[width=0.99\linewidth,keepaspectratio=true]{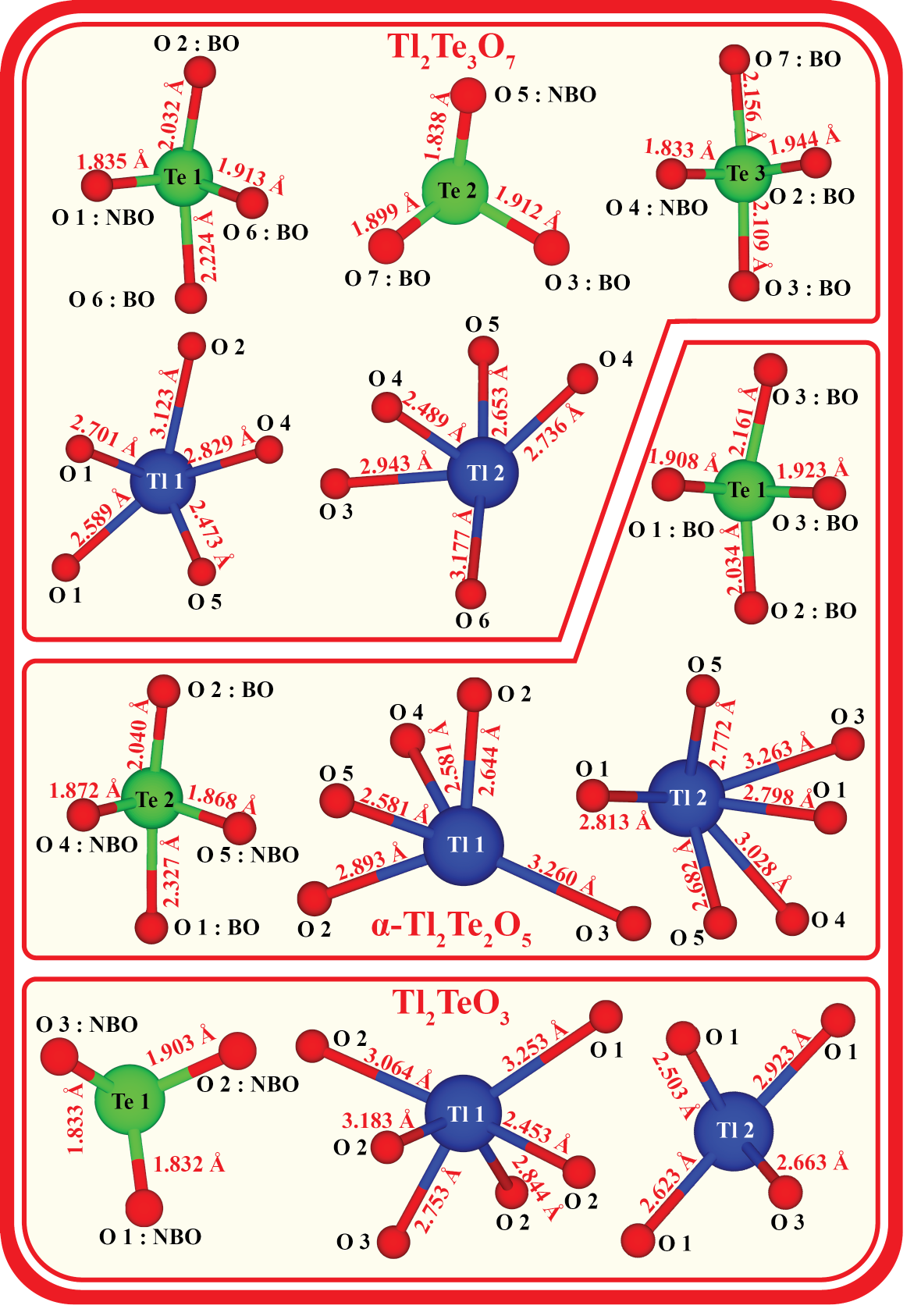}
    \caption{Illustration of various Tl and Te local environments in Tl$_2$Te$_3$O$_7$, $\alpha$-Tl$_2$Te$_2$O$_5$, and Tl$_2$TeO$_3$ crystalline polymorphs.}
    \label{label_figure7}
\end{figure}

Relying on the crystalline structures offers an additional reference point to support the trends observed in glasses for the calculated coordination numbers. We examined three Tl-Te-O crystalline structures, namely Tl$_2$Te$_3$O$_7$ \cite{jeansannetas1998crystal}, $\alpha$-Tl$_2$Te$_2$O$_5$ \cite{jeansannetas1997crystal} and Tl$_2$TeO$_3$ \cite{mirgorodsky2002dynamics}, which correspond to TlO$_{0.5}$ modifier concentrations of 40\%, 50\%, and 67\%, respectively. The computed values of n$_{\text{Te-O}}$ and n$_{\text{Tl-O}}$ for these crystals and the studied glasses are provided in Tables~\ref{tab4_coordination_various_glass} and \ref{tab4_Tl-O_Tl-NBO_coordination}. 
Interestingly, a trend of decreasing coordination numbers with increasing modifier concentration is also observed in the crystals, in agreement with the evolution observed in the studied glasses. The different Te$^{4+}$ and Tl$^{+}$ local environments present in the crystalline structures considered are illustrated in Figure~\ref{label_figure7}.

\begin{itemize}
\item For the Tl$_2$Te$_3$O$_7$ structure, three Te$^{4+}$ and two Tl$^{+}$ non-equivalent cations are present. Two Te$^{4+}$ are fourfold coordinated, the last one is threefold coordinated. This configuration yields an average n$_{\text{Te-O}}$ value of 3.67, closely matching the computed value of 3.72. The two distinct Tl$^{+}$ sites (Tl1 and Tl2) are fivefold coordinated, resulting in an average coordination number of 5, which closely matches the computed value of 5.03. The Tl1 atoms are bonded to one BO (O2) and four NBOs (O1, O1, O4, O5), while the Tl2 atoms are bonded to three NBOs (O4, O4, O5) and two BOs (O3, O6). This yields an average of 1.5 for n$_{\text{Tl-BO}}$ and 3.5 for n$_{\text{Tl-NBO}}$, values which align closely with the computed values of 1.46 and 3.58, respectively.

\item For the $\alpha$-Tl$_2$Te$_2$O$_5$ structure, two Te$^{4+}$ and two Tl$^{+}$ non-equivalent cations exist. All Te$^{4+}$ are fourfold coordinated, leading to a theoretical average n$_{\text{Te-O}}$ equal to 4. The computed value of 3.76 supports this expectation. The Tl sites present fivefold and sixfold coordination, leading to an average n$_{\text{Tl-O}}$ of 5.5, which is consistent with the computerd value of 5.36. The Tl1 atoms connect with three BOs (O2, O2, O3) and two NBOs (O4, O5), while the Tl2 atoms connect with three BOs (O1, O1, O3) and three NBOs (O4, O5, O5). These linkages yield average n$_{\text{Tl-BO}}$ and n$_{\text{Tl-NBO}}$ values of 3 and 2.5, respectively, which are not very far from the computed values of 2.25 and 3.10.

\item For the Tl$_2$TeO$_3$ structure, one Te$^{4+}$ and two Tl$^{+}$ cations are present. The Te$^{4+}$ cation is threefold coordinated, resulting in n$_{\text{Te}} = 3$, which aligns with the computed value of 3.01. The two Tl sites with sixfold and fourfold coordination give an average n$_{\text{Tl-O}}$ of 5, in reasonable agreement with the computed value of 4.67. Given that Te$^{4+}$ cations are only bonded to NBOs in this structure, the expected average n$_{\text{Tl-BO}}$ and n$_{\text{Tl-NBO}}$ values are 0 and 5, respectively. The calcultated values, 0.04 and 4.64, closely match the expected ones.
\end{itemize}

This comparison supports the validity of the calculated n$_{\text{Te-O}}$, n$_{\text{Tl-O}}$, n$_{\text{Tl-BO}}$ and n$_{\text{Tl-NBO}}$ values in $\rm {(TlO_{0.5})}_{\textit{y}}-{(TeO_2)}_{1-\textit{y}}$ binary glasses, reinforcing the consistency between crystal chemistry insights and our computational results.

\subsubsection{Atomic local environments}
\label{sec4_atomic_local_environment}

\begin{figure}[!htbp]
    \centering
    \includegraphics[width=0.99\linewidth,keepaspectratio=true]{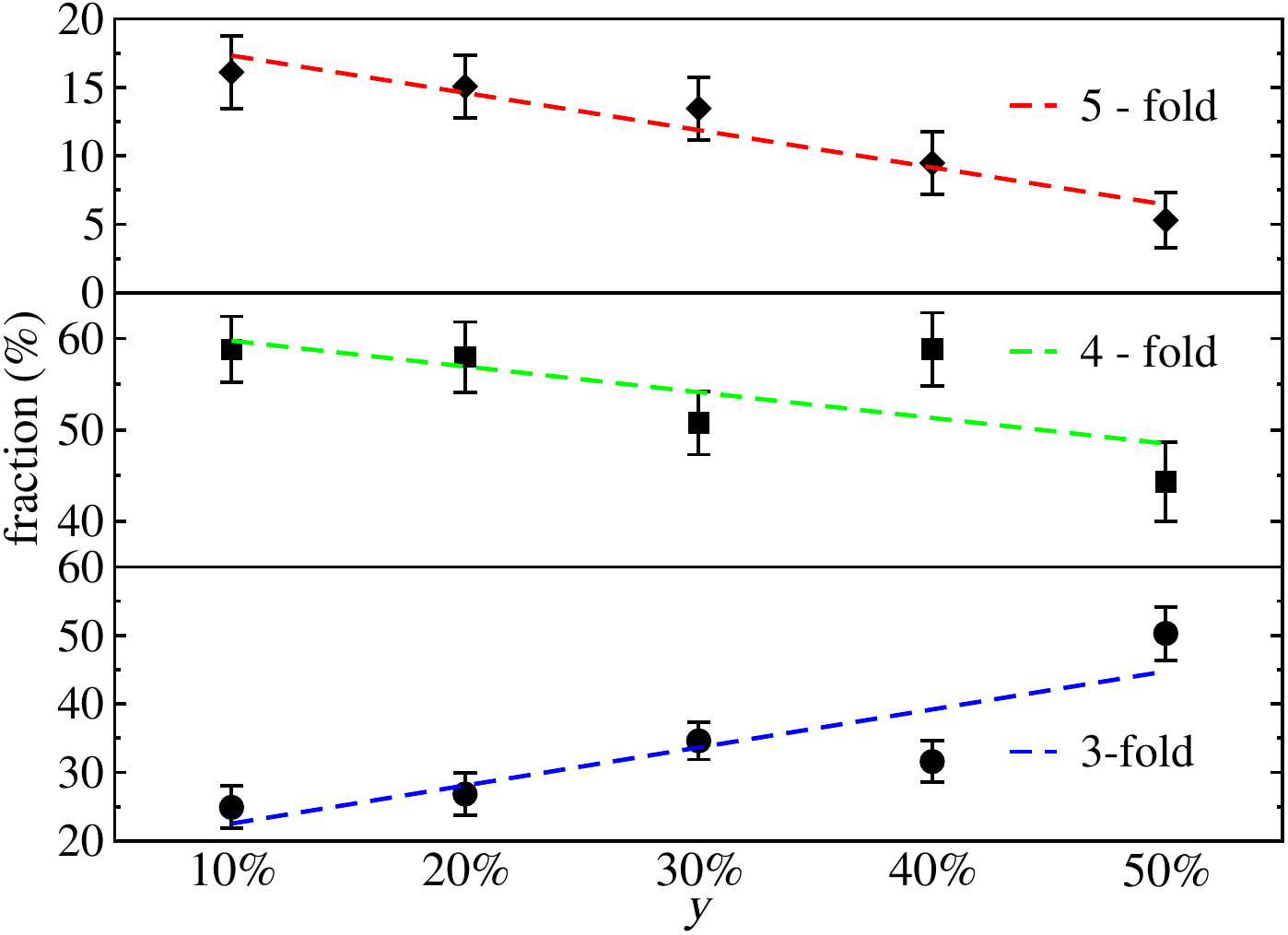}
    \caption{Fraction of $l$-fold coordinated Te$^{4+}$ cations in $\mathrm{(TlO_{0.5})}_{y}-\mathrm{(TeO_2)}_{1-y}$ glasses. Standard deviations are also provided. Units contributing less than 2\% are not shown.}
    \label{label_figure8}
\end{figure}

The decomposition of the local environment around Te$^{4+}$ cations is shown in Figure~\ref{label_figure8}. Only $l$-fold units contributing more than 2\% are included in the analysis. At low TlO$_{0.5}$ concentration, alongside the predominant 4-fold Te units, we also observe under-coordinated 3-fold and over-coordinated 5-fold Te units. Generally, as the concentration of TlO$_{0.5}$ increases, the fraction of TeO$_3$ units rises, while the total concentration of 4-fold and 5-fold units declines. This trend is due to the formation of NBOs.

We analyze the local environments around Te atoms through the $Q^n_m$ polyhedra, where $n$ represents BO and $m$ denotes the total number of oxide anions bonded to Te$^{4+}$. The $Q^n_m$ values for each glass composition are provided in Table~\ref{tab4_Te_qnm}. 
Focusing on the three-fold coordinated units, we observe a decline of fully connected TeO$_3$ units ($Q^3_3$) and the concurrent increase in partially connected $Q^1_3$ units with rising modifier concentration. Due to the opposing trends of increasing $Q^0_3$ and $Q^1_3$ populations and the decrease in $Q^3_3$ units, the $Q^2_3$ units reach a maximum at $y = 30\%$ modifier concentration and remain relatively constant at approximately 12$-$14\% for all concentrations.
At TlO$_{0.5}$ concentration of $y = 50\%$, we observe a significant presence of isolated TeO$_3$ units ($Q^0_3$), indicating that Tl promotes full structural depolymerization by replacing Te$-$BO$-$Te bridges with Te$-$NBO$-$Tl linkages. This finding highlights the role of thallium in breaking the connectivity of Te polyhedra. 

\begin{table}[!htbp]
    \centering
    \caption{Q$^n_m$ values (in percentage) evaluated for Te environments with r$_{\text{cut-off}}$ = 2.46 {\AA}. Values less than 2\% are not shown.}
    \begin{tabular}{c | c | c | c | c | c}
        \hline \hline
        Q$^n_m$ & $y$ = 10\% & $y$ = 20\% & $y$ = 30\% & $y$ = 40\% & $y$ = 50\% \\
        \hline \hline
        Q$_3^0$ & -          & -          & -          & -          & 8.22       \\
        Q$_3^1$ & -          & 3.1        & 8.47       & 11.44      & 25.99      \\
        Q$_3^2$ & 11.24      & 12.21      & 18.06      & 15.62      & 14.58      \\
        Q$_3^3$ & 12.57      & 11.45      & 7.61       & 3.39       & -          \\
        \hline
        Q$_4^2$ & -          & 2.45       & 2.9        & 8.19       & 10.61      \\
        Q$_4^3$ & 20.18      & 22.4       & 26.88      & 33.05      & 26.06      \\
        Q$_4^4$ & 37.43      & 33.06      & 20.86      & 17.06      & 6.17       \\
        \hline
        Q$_5^4$ & 2.97       & 2.49       & 2.67       & 3.84       & 4.2        \\
        Q$_5^5$ & 13.11      & 12.5       & 10.71      & 5.38       & -          \\
        \hline \hline
    \end{tabular}
    \label{tab4_Te_qnm}
\end{table}

Focusing on the four-fold coordinated Te$^{4+}$ cations, similar trends to those observed in the case of three-fold coordinated units are observed where the population of $Q^2_4$ units increase at the expanse of the $Q^4_4$ units. Due to this opposite trend, the $Q^3_4$ units display an evolving pattern up to 40\% modifier concentration before decreasing. In the case of over-coordinated Te polyhedra, we observe a decreasing trend in the $Q^5_5$ population as the modifier oxide concentration increases. The $Q^4_5$ units exhibit a minimal presence, remaining roughly constant at approximately 3\% across all concentrations.

\begin{figure}[!htbp]
    \centering
    \includegraphics[width=0.99\linewidth,keepaspectratio=true]{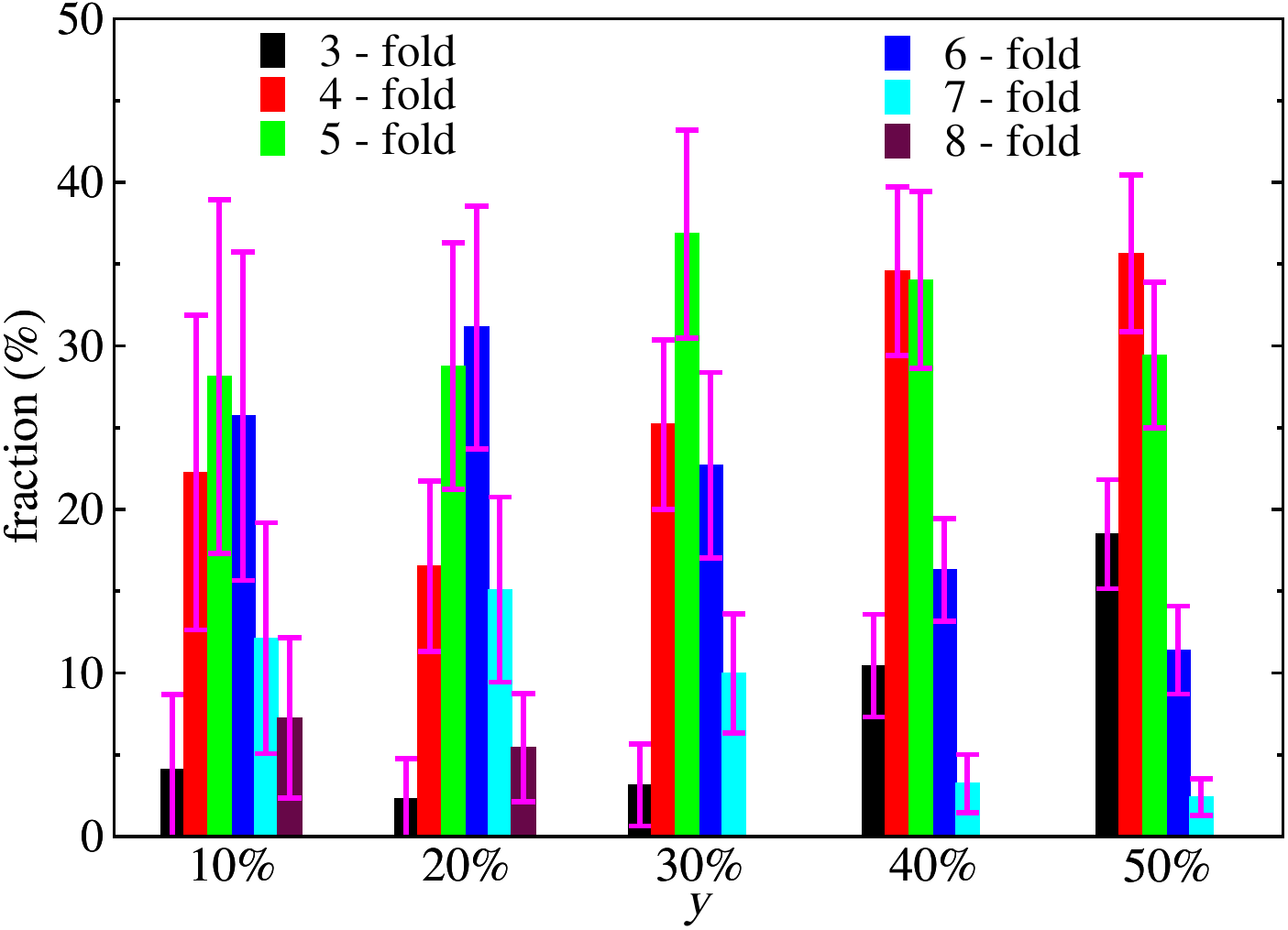}
    \caption{Fraction of $l$-fold coordinated Tl$^{+}$ cations in $\rm {(TlO_{0.5})}_{\textit{y}}-{(TeO_2)}_{1-\textit{y}}$  glasses. Values less than 2\% are ignored.}
    \label{label_figure9}
\end{figure}

To investigate the local environment of Tl$^{+}$ cations, we plot the $l$-fold fractions  across different modifier concentrations using a cut-off radius of 3.26 {\AA} (see Figure~\ref{label_figure9}). The Tl coordination numbers range between 3 and 8 with the 3-fold and 4-fold Tl coordinations becoming more prominent for the highest modifier concentration. 

\begin{figure}[!htbp]
    \centering
    \includegraphics[width=0.9\linewidth,keepaspectratio=true]{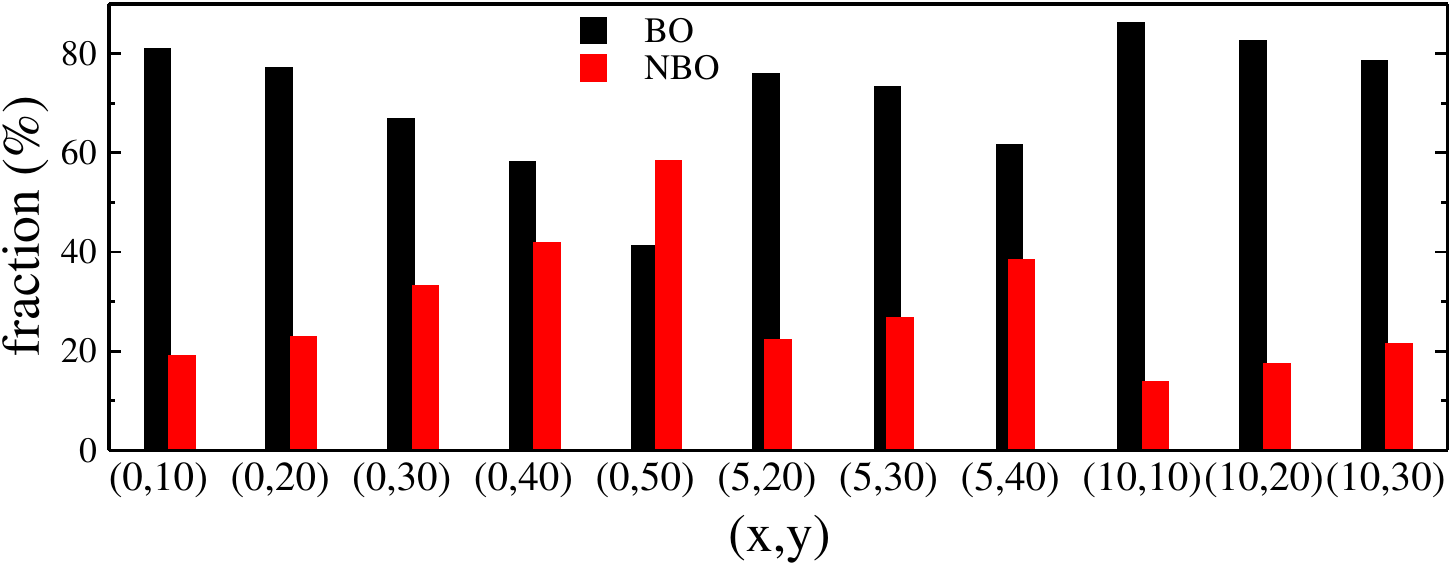}
    \caption{Fractoin of BO and NBO in $\rm {(TlO_{0.5})}_{\textit{y}}-{(TeO_2)}_{1-\textit{y}}$ and $\rm{(TiO_{2})}_{\textit{x}}-{(TlO_{0.5})}_{\textit{y}}-{(TeO_2)}_{1-\textit{x}-\textit{y}}$ glasses.}
    \label{label_figure10}
\end{figure}

Finally, the evolution of BO and NBO fractions (with respect to Te atoms) across different modifier concentration is plotted in Figure~\ref{label_figure10}). 
We find a monotonous increase of the BO population at the expanses of the NBO population with a population crossover occurring at 50\% modifier concentration and leading to more NBOs in the glass than BOs.

Taken together, the decrease of the Te coordination number with its related shortening average Te-O bond length, the increase of NBO population reflected by a high population of Q$^n_m$ with $n < m$ and the spatial proximity of Tl$^{+}$ cations with NBO show that the addition of TlO$_2$ leads to a strong structural depolymerization of the parent TeO$_2$ network. 

\subsubsection{Ring statistics}

\begin{figure}[!htbp]
    \centering
    \includegraphics[width=0.99\linewidth]{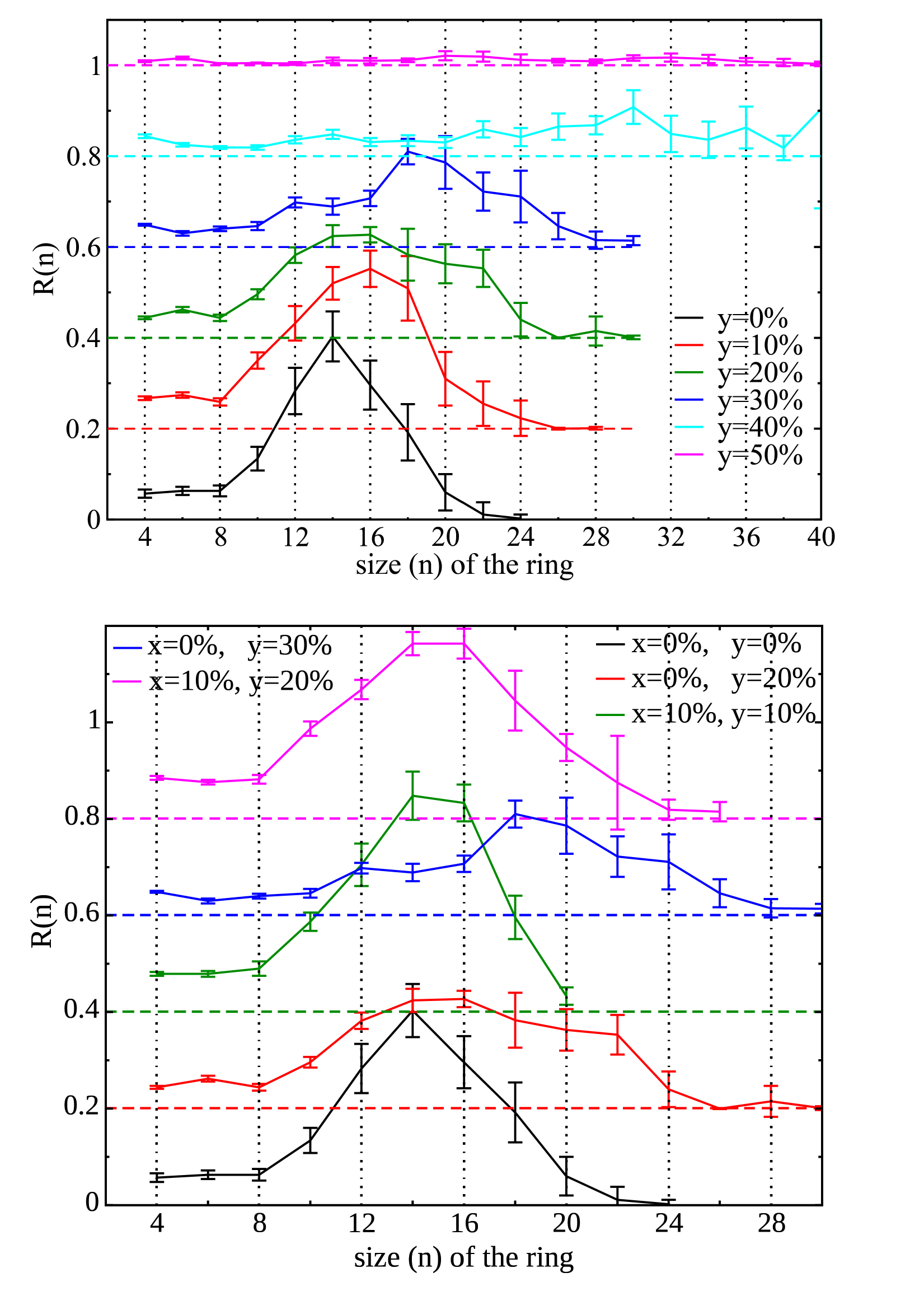}
    \caption{Rings size distribution profile in $\rm {(TlO_{0.5})}_{\textit{y}}-{(TeO_2)}_{1-\textit{y}}$ binary glasses (top panel) compared to $\rm {(TiO_{2})}_{\textit{x}}-{(TlO_{0.5})}_{\textit{y}}-{(TeO_2)}_{1-\textit{x}-\textit{y}} $ ternary glasses (bottom panel). For clarity, a vertical shift along with horizontal dashed line representing the zeros of various concentration curves are provided.}
    \label{label_figure11}
\end{figure}

\begin{figure}[!htbp]
    \centering
    \includegraphics[width=0.99\linewidth]{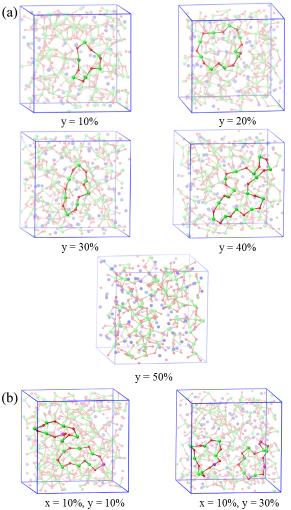}
    \caption{Representation of rings in $\rm {(TlO_{0.5})}_{\textit{y}}-{(TeO_2)}_{1-\textit{y}}$ binary (a) and $\rm{(TiO_{2})}_{\textit{x}}-{(TlO_{0.5})}_{\textit{y}}-{(TeO_2)}_{1-\textit{x}-\textit{y}}$ ternary (b) glasses. Color code of atoms, red: oxygen, green: tellurium, blue: thallium. }
    \label{label_figure12}
\end{figure}

Ring statistics provide complementary information and allow exploring the connectivity of amorphous materials at a larger scale than the first or second coordination shells. We resort to the Rigorous Investigation of Networks Generated using Simulation (RINGS) software package to identify network connectivity profiles \cite{roux_ring_2010} between various atomic species using the King/Franzblau \cite{king1967ring, franzblau1991computation} shortest paths criterion and using the established distance cutoffs for Te-O and Tl-O. 
In both systems, we search for rings formed with maximum size of 40 atoms. 
The numbers of rings, R(n), contain n ions normalized to the total number of ions. For all the systems the results were averaged on 100 configurations extracted from the last 10 ps of MD trajectories at T = 300 K. The obtained distribution of rings sizes is displayed in Figure \ref{label_figure11}. Figure \ref{label_figure12}, showcase snapshots containing representative  rings of the largest size in binary glasses with various compositions.

We find that the peak of R(n) for pure TeO$_2$ glass is centered around n = 14. The increase of $\mathrm{TlO_{0.5}}$ concentration leads to a shift of R(n) distribution towards larger values of n and hence, reflecting an increase of the average rings size, and its intensity flattens, reflecting a wider distribution of rings. This evolution can be understood on the basis of the structural depolymerization of the glassy network as it leads to replace -Te-O-Te- bridges by $-$Te$=$O$^{-}$$\cdots$Tl$^{+}$$-$ linkages and hence opening smaller rings that might combine to form the larger rings. 
In the specific case of $y$ = 50\% system, the occurrence of rings compatible with the simulation box size is almost zero, reflecting that the $-$O$-$Te$-$O$-$Te$-$O$-$ rings being completely depolymerized by Tl$^{+}$ (see Figure \ref{label_figure12}). 

\subsubsection{Chemical bonds characterisation}

The classification of chemical bonding in terms of ionic and covalent character 
brings further details about the glass network connectivity. To access this property, we compute the electronic localization function (ELF) and analyze the degree of ionicity or covalency of bonding in the glassy matrix. The ELF is scaled between 0 and 1, ELF very close to 1 marks strongly localised pairs (lone pairs, covalent bonds), ELF = 0.5 is the uniform-electron-gas reference and, in inhomogeneous systems, values near this often signal metallic/delocalised bonding, ELF approaching 0 occurs in nodal/interbasin with very low-localisation regions. As such, ELF provides a spatial map of chemical bonding, distinguishing covalent, ionic, and metallic character based on the degree of electron localization.

\begin{figure}[!htbp]
     \centering
     \includegraphics[width=0.99\linewidth,keepaspectratio=true]{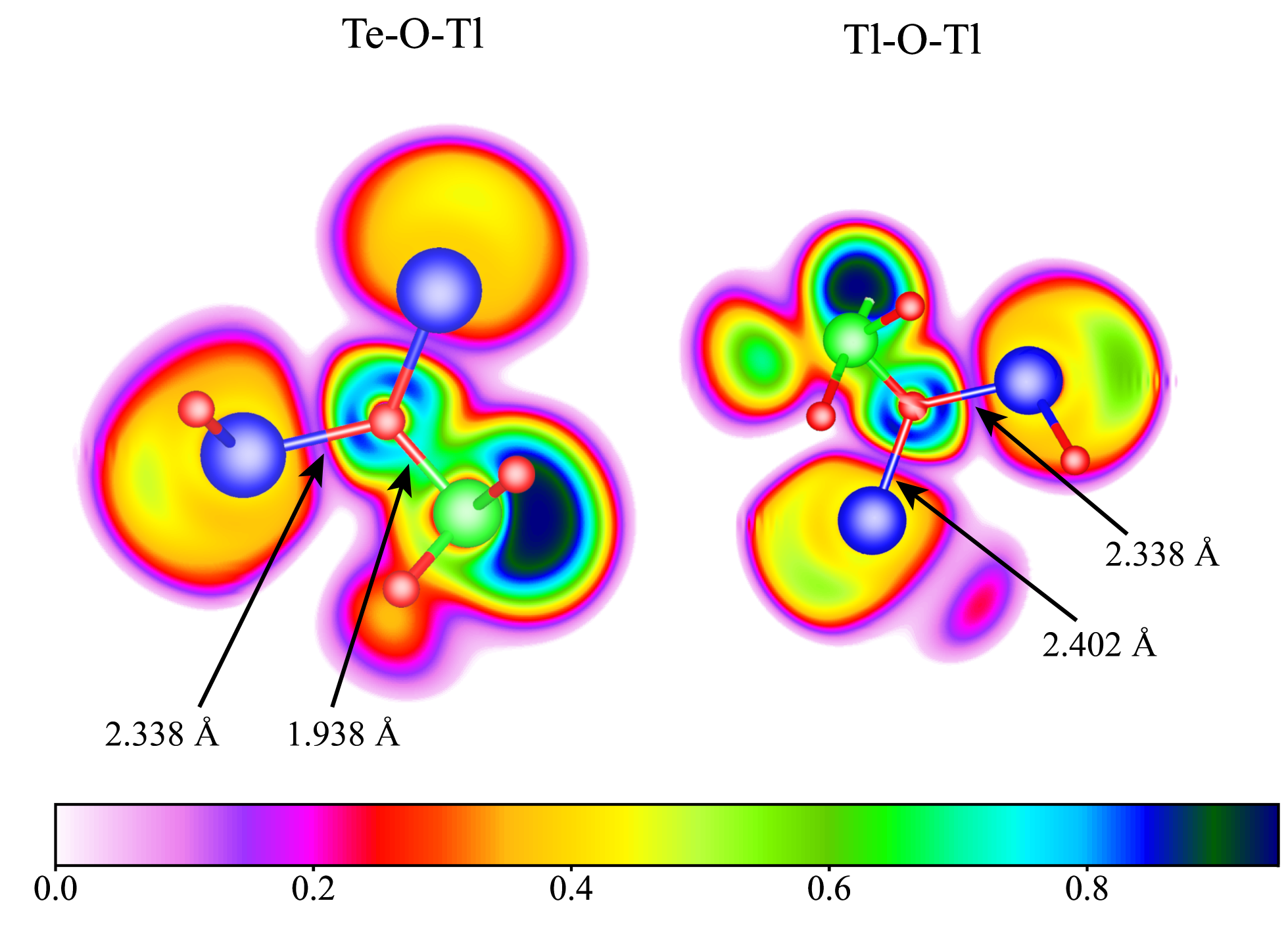}
     \caption{Electron localization function representation described for Tl-O-Tl (right) bridge and Te-O-Tl (left) bridge. Red, green and blue sphere show oxygen, tellurium, and thallium atoms. Arrows indicate bonds in the projection plane.}
    \label{label_figure13}
\end{figure}

We have selected one structural motif from the glassy structure of $y$ = 30\% of $\rm {(TlO_{0.5})}_{\textit{y}}-{(TeO_2)}_{1-\textit{y}}$ binary glass and passivated oxygen atoms with hydrogen atoms in order to maintain charge neutrality. Thereafter, we have relaxed the atomic positions of hydrogen atoms before calculating ELF. We chose the PBE0 hybrid functional to compute ELF since it has better performance for localizing the electron density than GGA functionals. Figure \ref{label_figure13} shows the two-dimensional contour plots of the ELF in a plane spanned by Tl-O(NBO)-Tl (Figure \ref{label_figure13}, left) and Te-O(NBO)-Tl (Figure \ref{label_figure13}, right) bridge. One can see from this figure that the Te-O-Tl bridge has a strong ELF value of about 0.6 in the vicinity of Te and the bonding O, whereas the Tl and O linkage exhibits its minimum at less than 0.2, indicating the highly ionic nature of the Tl-O bond when compared to the more covalent Te-O bond. When it comes to the Tl-O-Tl (Figure \ref{label_figure13}, left) bridge, we can see that both Tl-O linkages exhibit a fairly symmetric gradient in ELF values to the bond's center, with a minima smaller than 0.2 signifying a bond's identical ionic character. 

It is noteworthy that Te and Tl cations contain 5$s^2$ and 6$s^2$ lone pair respectively, however, one can recognize from the Figure \ref{label_figure13} that the probability distribution of electron pair shows a strong electronic localization around Te$^{4+}$ cation (dark blue) in comparison with the highly diffused electronic localization for Tl$^{+}$ lone pair (distorted circular yellow region). This observation gives further hints about the increasing accuracy of the calculated total X-ray scattering PDF in Figure \ref{label_figure1} with increasing of the modifier concentration. 
As the electronic localization of Tl$^{+}$ lone pair show a diffusive nature, PBE functional is able of describe it without the need of higher levels of theory, thereby leading to a better description of the network connectivity  for larger modifier concentrations. Conversely, at lower modifier concentrations, Te$^{4+}$ becomes a key part in the glassy structure and due to the strong electronic localization of its lone pair, a higher level of theory, such as hybrid functionals, is necessary to accurately describe the structure and reproduce experimental x-ray PDFs \cite{raghvender2022structure}.

\subsection{Structural effects of the addition of TiO$_2$}

In this section, we describe the effect of the addition of TiO$_2$ into the $\rm {(TlO_{0.5})}_{\textit{y}}-{(TeO_2)}_{1-\textit{y}}$ glasses. The partial PDFs are shown in Figure S3 and the first peak position are given in Table S1. 

At constant $\mathrm{TlO_{0.5}}$ concentration and increasing TiO$_2$ concentration, while we observe no significant shift in the peak position of g$_\text{Te-O}$($r$), there is a tiny shift of the first peak in g$_{\text{O-O}}(r)$ to smaller distances. Such changes are similar to what is observed in crystalline phases, for example between $\mathrm{TeO_{2}}$ and TiTe$_3$O$_3$ \cite{bindi2003crystal}, where the addition of TiO$_2$ strongly shorten the O-O distances in TeO$_m$ units but does not affect the Te-O average bond length.

The cationic pair correlations involving Ti$^{4+}$ show stronger statistical fluctuations due to the limited fraction of TiO$_2$ modifier and no clear trend can be extracted. In the particular case of g$_{\text{Ti-Ti}}(r)$ partial PDFs, we find two peaks in the range of 3 to 4 {\AA} that can be attributed to Ti$^{4+}$ cations connected via edge or corner sharing. 

\begin{table*}[!htbp]
    \centering
        \caption{Coordination numbers n$_{\text{Te-O}}$, n$_\text{Ti-O}$ and  n$_{\text{Tl-O}}$ = n$_\text{Tl-BO}$ + n$_\text{Tl-NBO}$ for all the considered modifier concentration in $\rm  {(TiO_{2})}_{\textit{x}}-{(TlO_{0.5})}_{\textit{y}}-{(TeO_2)}_{1-\textit{x}-\textit{y}} $ ternary glass systems.}
    \begin{tabular}{ c | c | c |c | c | c}
        \hline \hline
                  Compositions &  n$_{\text{Te-O}}$ &n$_\text{Ti-O}$ & n$_\text{Tl-BO}$  &  n$_\text{Tl-NBO}$ &  n$_\text{Tl-O}$  \\
        \hline \hline
        $x$ = 5\%, $y$ = 20\% & 3.86 $\pm$ 0.03 & 5.35 $\pm$ 0.06 & 3.69 $\pm$ 0.15       &  1.80 $\pm$ 0.12       &   5.50 $\pm$ 0.11 \\
        $x$ = 5\%, $y$ = 30\% & 3.87 $\pm$ 0.04 & 5.06 $\pm$ 0.05 & 2.74 $\pm$ 0.11       &  2.20 $\pm$ 0.09       &   4.94 $\pm$ 0.10\\
        $x$ = 5\%, $y$ = 40\% & 3.74 $\pm$ 0.05 & 5.26 $\pm$ 0.11 & 2.00 $\pm$ 0.09       &  2.62 $\pm$ 0.09       &   4.62 $\pm$ 0.08\\ 
        $x$ = 10\%, $y$ = 10\% & 3.91 $\pm$ 0.04 & 5.21 $\pm$ 0.07 & 4.18 $\pm$ 0.20       &  1.45 $\pm$ 0.16       &   5.63 $\pm$ 0.18\\
        $x$ = 10\%, $y$ = 20\% & 3.95 $\pm$ 0.03 & 5.28 $\pm$ 0.08 & 3.93 $\pm$ 0.15       &  1.69 $\pm$ 0.11       &   5.62 $\pm$ 0.14\\
        $x$ = 10\%, $y$ = 30\% & 3.93 $\pm$ 0.04 & 5.26 $\pm$ 0.06 & 3.37 $\pm$ 0.12       &  1.73 $\pm$ 0.10       &   5.10 $\pm$ 0.09\\
    \hline \hline
    \end{tabular}
    \label{tab5_TeTi_coordination_number}
\end{table*}

The coordination numbers are reported in Table \ref{tab5_TeTi_coordination_number}. In the case of n$_{\text{Te-O}}$ we used r$_\text{cut-off}$ = 2.48 {\AA} obtained using the MLWF formalism, a very close value to that obtained in the case of the binary system. As for n$_{\text{Ti-O}}$, we used the first minimum position in the corresponding partial PDF located at 2.50 {\AA} (see Figure S3). At these cutoff values, the running coordination numbers n$_{\text{Te-O}}$(r) and n$_{\text{Ti-O}}$(r) reach a plateau region as shown in Figure \ref{label_figure4}.
At $x= $ 5\% $\mathrm{TiO_{2}}$ concentration, the n$_{\text{Te-O}}$ values and there evolution with $\mathrm{TlO_{0.5}}$ content are similar to those obtained in the corresponding binary systems. However, at $x= $ 10\% $\mathrm{TiO_{2}}$ concentration, we observe a substantial increase in the n$_{\text{Te-O}}$ to values close to 4 with no effect of the $\mathrm{TlO_{0.5}}$ content.  
This behaviour can be understood by the network forming nature of TiO$_2$ which tends to counteract the depolymerization induced by $\mathrm{TlO_{0.5}}$. This effect, is only noticeable at $\mathrm{TiO_{2}}$ concentration of $x= $ 10\%. The corresponding fractions of Te Q$^n_m$ units are reported in Table \ref{tab5_Te_qnm}. 
It is interesting to note that, with increasing $\mathrm{TiO_{2}}$ content while keeping the $\mathrm{TlO_{0.5}}$ content constant, fully connected Q$^4_4$ and Q$^5_5$ structural units are favored in agreement with the above observations. 

\begin{table}[!htbp]
    \centering
         \caption{Q$^n_m$ values (in percentage) evaluated for Te environments in  $\rm  {(TiO_{2})}_{\textit{x}}-{(TlO_{0.5})}_{\textit{y}}-{(TeO_2)}_{1-\textit{x}-\textit{y}}$ glasses computed using r$_{\text{cut-off}}$ = 2.48 {\AA} and the Wannier formalism constraints. Values less than 2\% are not shown.}
    \begin{footnotesize}
     \begin{tabular}{c | c | c | c | c | c | c}
        \hline \hline 
                    Q$^n_m$   & $x$ = 5\% &  $x$ = 5\% & $x$ = 5\% &  $x$ = 10\%  & $x$ = 10\% & $x$ = 10\%   \\
                     units    & $y$ = 20\% &  $y$ = 30\% & $y$ = 40\% & $y$ = 10\% &  $y$ = 20\% & $y$ = 30\%  \\
          \hline \hline
            Q$_3^1$   &   3.21   &   3.59  &    10.33    &    -       &     -      &  2.58    \\ 
            Q$_3^2$   &  13.71   &  13.82  &   18.11     &  10.63     &   10.21    & 12.65    \\
            Q$_3^3$   &  11.63   &  11.05  &    6.71     &  14.26     &    9.16    &  8.3     \\
            \hline 
            Q$_4^2$   &    2.65  &   3.85  &   10.29     &    -       &    2.04    & 3.03     \\
            Q$_4^3$   &   20.69  &  28.02  &   28.19     &  14.99     &   20.32    & 22.94    \\
            Q$_4^4$   &   33.33  &  24.25  &   16.3      &  41.42     &   41.13    & 33.26    \\
            \hline
            Q$_5^4$   &   2.53   &  4.49   &   4.83      &    -       &     -      &  3.32    \\
            Q$_5^5$   &   11.9  &   10.7   &   4.48      &  15.79     &   14.04    &  13.8    \\
        \hline \hline
     \end{tabular}
     \end{footnotesize}
     \label{tab5_Te_qnm}
\end{table}

The average coordination number of thallium, n$_{\text{Tl-O}}$, increases as a function of increasing TiO$_2$ content from $x= $ 5\% to $x= $ 10\% while keeping the TlO$_{0.5}$ amount constant. This behaviour corresponds to an opposite evolution of the Tl-NBO and Tl-BO connections, with the latter showing a faster increase at $x= $ 10\%.
This rise in n$_\text{Tl-BO}$ can be attributed to a rise of the BO population with increasing TiO$_2$. This is clearly observed in Figure \ref{label_figure10} and is in
good agreement with the analysis of the coordination number (Table \ref{tab5_TeTi_coordination_number}).  

\begin{figure}[!htbp]
    \centering
        \includegraphics[width=0.9\linewidth]{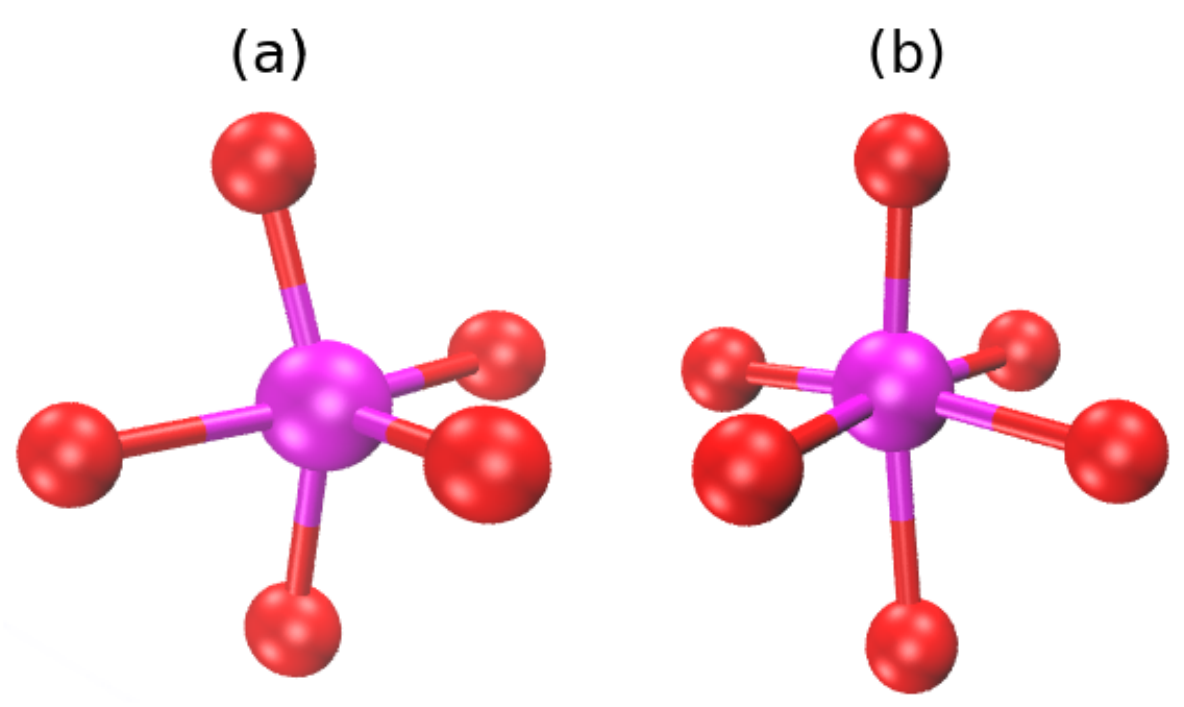}
        \caption{Pentahedral TiO$_5$ and octahedral TiO$_6$ units in  $\rm  {(TiO_{2})}_{\textit{x}}-{(TlO_{0.5})}_{\textit{y}}-{(TeO_2)}_{1-\textit{x}-\textit{y}} $ glasses. Red sphere: oxygen, magenta sphere: titanium.}
    \label{label_figure14}
\end{figure}

Coming to the n$_\text{Ti-O}$ coordination number, we find that it is 
around 5.23 in average (see Table \ref{tab5_TeTi_coordination_number}) and show a little dependence with the glass composition. Similar n$_\text{Ti-O}$ value has been observed experimentally in liquid \cite{alderman2014structure}, sputtered and powder amorphous \cite{petkov1998atomic} TiO$_2$, 5.0, 5.4 and 5.6, respectively.
This value reflects the the existence of a majority of 5-fold (pentahedral TiO$_5$, see Figure \ref{label_figure14}(a)]) together with 4-fold (tetrahedral TiO$_4$) and 6-fold (octahedral TiO$_6$, see Figure \ref{label_figure14}(b)]) environments in the glass.

Finally, Figure \ref{label_figure11} illustrates the evolution of the rings distributions. Interestingly, we find that substituting TlO$_{0.5}$ with TiO$_2$ in $\rm{(TiO_{2})}_{\textit{x}}-{(TlO_{0.5})}_{\textit{y}}-{(TeO_2)}_{1-\textit{x}-\textit{y}}$ ternary glasses, while keeping the molar concentration of TeO$_2$ units constant, shifts the peak of the distribution towards lower values of n, indicating that TiO$_2$ helps in repolymerizing the structure and thus making rings of smaller size. Representative rings forming units in  (TiO$_2$)$_{0.1}$-(TlO$_{0.5}$)$_{0.1}$-(TeO$_2$)$_{0.8}$ and (TiO$_2$)$_{0.1}$-(TlO$_{0.5}$)$_{0.3}$-(TeO$_2$)$_{0.6}$ ternary glasses are depicted in Figure \ref{label_figure12} which illustrates the fact that Ti$^{4+}$ participates in forming the rings by substituting $-$O$-$Te$-$O$-$Te$-$O$-$ chains by $-$O$-$Te$-$O$-$Ti$-$O$-$ chains. This observation indicates that Ti$^{4+}$ ions promote the polymerization of the glassy network as previously explained.

\subsection{Vibrational and non-linear properties}

\subsubsection{Raman spectroscopy}
\begin{figure}[!htbp]
    \centering
    \includegraphics[width=0.99\linewidth]{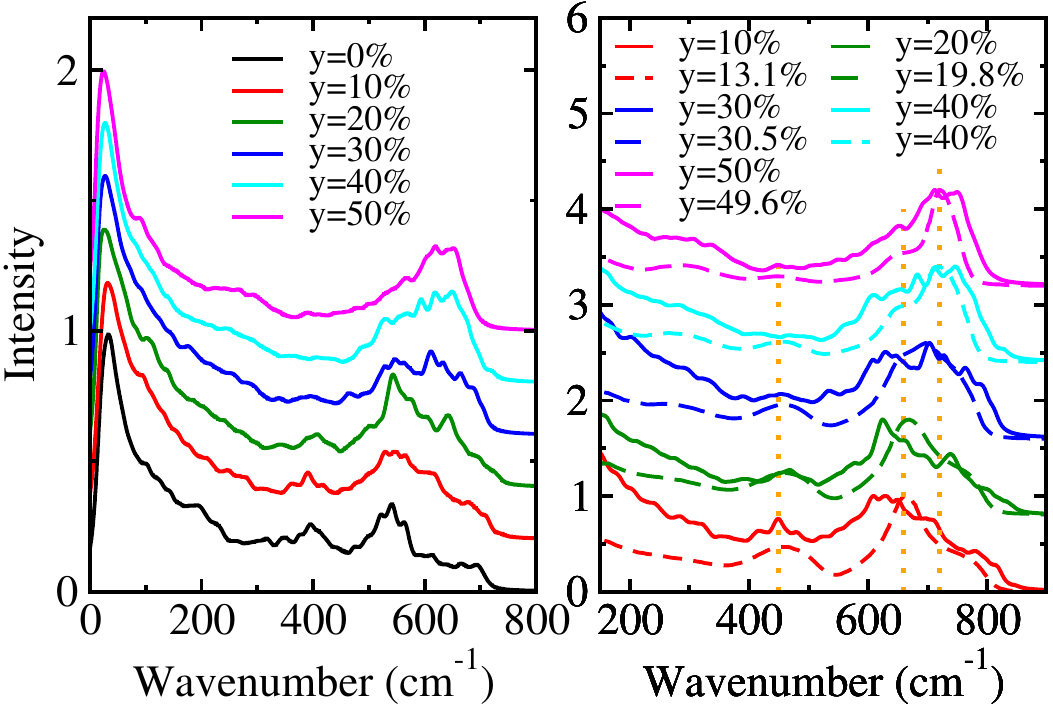}
    \caption{Computed Raman spectra (solid lines) of $\mathrm{(TlO_{0.5})}_{y}-\mathrm{(TeO_2)}_{1-y}$ binary glasses as obtained (left side) and with 15\% blue shift (right side), compared to experimental data\cite{udovic2009formation} (dashed lines). 
    A vertical shift was applied for clarity and vertical dotted lines are provided at the position of the main Raman bands.}
    \label{label_figure15}
\end{figure}

\begin{figure}[!htbp]
    \centering
        \includegraphics[width=0.99\linewidth]{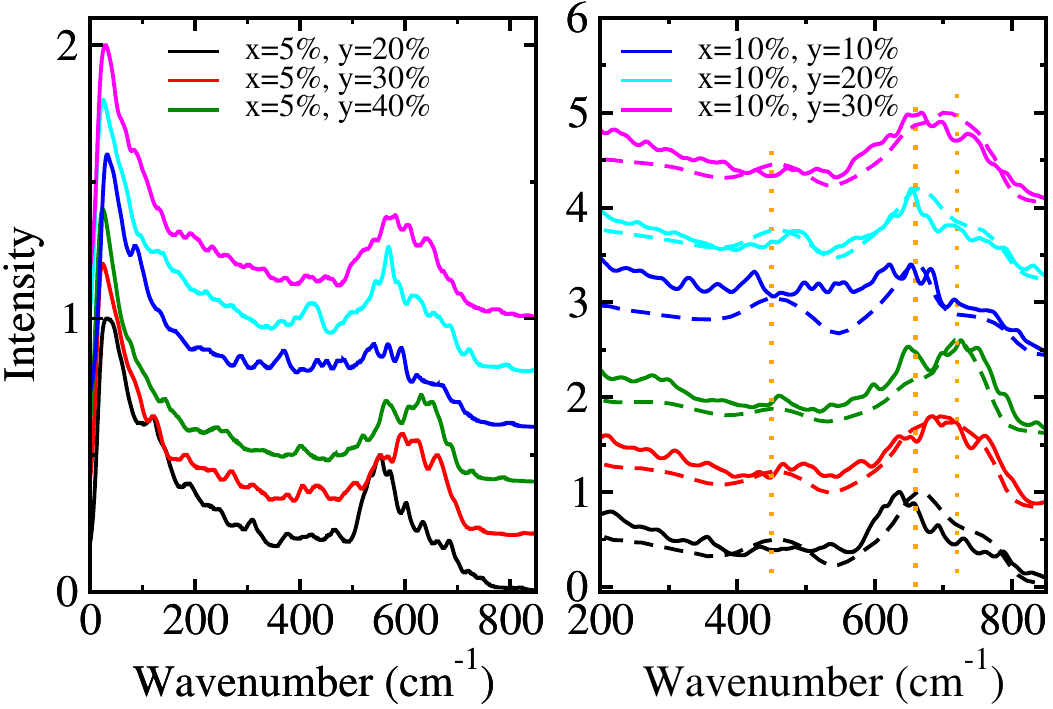}
        \caption{Computed Raman spectra (solid lines) of $\rm  {(TiO_{2})}_{\textit{x}}-{(TlO_{0.5})}_{\textit{y}}-{(TeO_2)}_{1-\textit{x}-\textit{y}}$ ternary glasses as obtained (left side) and with 15\% blue shift (right side), compared to experimental data\cite{udovic2009formation} (dashed lines). 
    A vertical shift was applied for clarity and vertical dotted lines are provided at the position of the main Raman bands.}
    \label{label_figure16}
\end{figure}

Raman spectra were computed on the fully periodic amorphous models of both $\mathrm{(TlO_{0.5})}_{y}-\mathrm{(TeO_2)}_{1-y}$ binary and $\rm  {(TiO_{2})}_{\textit{x}}-{(TlO_{0.5})}_{\textit{y}}-{(TeO_2)}_{1-\textit{x}-\textit{y}} $ ternary glasses by resorting to the scheme described in our recent work \cite{roginskii2023ab}. Figure~\ref{label_figure15} and Figure~\ref{label_figure16} show the evolution of Raman spectra for various compositions of the binary and ternary glasses, respectively. For each composition, the simulated Raman spectra at T = 300 K is compared to available experimental data. To account for the overestimation of Te$-$O bond lengths in our models, when comparing to experiments, a blueshift of 15\% was applied to the wavenumber for each computed spectra following Ref. (\citenum{roginskii2023ab}). In all spectra, the most intense peak, centered around 30 cm$^{-1}$, is attributed to the Boson peak contribution, which is a characteristic of low-frequency vibrational modes in glassy systems. Due to the lack of experimental data on the low-frequency region, particularly for the ternary glasses, the discussion will focus on the 200$-$850 cm$^{-1}$ frequency range of the Raman spectra.

Our models successfully reproduce all the major bands observed in the experimental spectra of binary glasses, as well as their evolution as a function of the system composition (Figure~\ref{label_figure15}). The spectrum of pure TeO$_2$ glass ($y=0$) show two main features: a broad peak in the middle frequency range (400$-$500 cm$^{-1}$) and a dominant peak in the higher frequency range at 660 cm$^{-1}$. The addition of the thallium oxide modifier has two main effects on the spectra shape: the broad peak in the middle frequency range gradually disappears reflecting the drastic descrease in the population of BO. The dominant peak in higher frequency range shifts from 660 cm$^{-1}$ to 725 cm$^{-1}$. This behaviour was already observed in previous studies \cite{sekiya, udovic2009formation, NOGUERA2004981} and was attributed to the structural depolymerization of the glass where TeO$_4$ units transforms into TeO$_3$ one with the addition of TlO$_{0.5}$ modifier oxide, as discussed previously. 

As for the ternary glasses, the simulated Raman spectra show an overall good agreement with available experimental data (see Figure \ref{label_figure16}). 
Compared to binary glasses, we find that the positions and the intensities of the peaks around 460 and 660 cm$^{-1}$ did not significantly change by replacing TeO$_2$ with TiO$_2$ at constant TlO$_{0.5}$ fraction. This suggests that Ti behaves similarly to Te and maintains the polymerization of the glassy framework. In practice, titanium-oxide substitute the -O-Te-O$\cdots$Te-O- linkages by stronger -O-Te-O-Ti-O-Te-O- bridges, where both Te-O and Ti-O features very close bond lengths. 
Moreover, increasing TiO$_2$ modifier oxide from $x = $5\% to $x = $10\% in ternary glasses at constant TlO$_{0.5}$ concentration shows that the broad peak becomes slightly more pronounced, which indicates that Ti$^{4+}$ in Te-O-Ti bridges has better capability in polymerizing the glassy network than Te$^{4+}$ in Te-O-Te bridges. This pronounced effect allows one to deduce that no structural depolymerization of the glassy network occurs as the peak around $\sim$ 660 cm$^{-1}$ shows a limited change. These observations depict that, incorporation of Ti$^{4+}$ does not lead to formation of NBO unlike Tl$^{+}$.

At constant TiO$_2$ and increasing TlO$_{0.5}$ fractions a similar evolution of the Raman sepctra shape is found as that observed in the case of binary $\rm {(TlO_{0.5})}_{\textit{y}}-{(TeO_2)}_{1-\textit{y}}$ glasses.

\subsubsection{Non-linear optical properties}
The ultimate goal of investigating \ch{TeO2} modified glasses is to understand how the chemical composition and the structural properties of these materials influence their 
nonlinear optical properties, particularly their third-order nonlinear susceptibilities, $\chi^{(3)}$. In addition, it is of interest to assess the ability of the established glassy models to reproduce the available experimental $\chi^{(3)}$ data \cite{duclere2018third}. To this end, we resort to our recently developed computational scheme to compute the NLO properties on periodic glassy models. This scheme is based on a combination of linear response theory and finite difference method, and was applied and validated in the case pure TeO$_2$ glass \cite{roginskii2024nonlinear}. 
For detailed methodological and technical details, we refer the reader to Ref. (\citenum{roginskii2024nonlinear}).

In amorphous systems, the mean values of $\chi^{(1)}$, diagonal $\chi^{(3)}_d$, and non-diagonal $\chi^{(3)}_{nd}$ components are calculated as the arithmetic averages of their respective tensor elements.  The overall mean value of the third-order nonlinear susceptibility, $\langle\chi^{(3)}\rangle$, is determined using the following expression:

\begin{equation}
    \langle\chi^{(3)}\rangle=\frac{1}{5}\left ( \chi^{(3)}_{xxxx} + \chi^{(3)}_{yyyy} + \chi^{(3)}_{zzzz} + 2\chi^{(3)}_{xxyy} + 2\chi^{(3)}_{xxzz} + 2\chi^{(3)}_{yyzz}      \right )
    \label{Eq:chi3mean}
\end{equation}

For each considered glass composition, statistical averages are performed over 10 snapshots extracted from the MD trajectories at T = 300K.
Experimentally, non-linear susceptibility measurements are often performed relative to a reference material. For instance, NLO properties of paratellurite single crystals are typically measured against a fused silica reference glass \cite{duclere2018third}. In our study, we adopt a similar approach by considering a reference \ch{SiO2} model on top of which we evaluate the NLOP. Our previous work demonstrated that in the case of pure TeO$_2$ glasses, extending the simulation size from approximately 500 atoms to 2000 atoms altered the calculated $\chi^{(3)}$ ratio between TeO$_2$ and SiO$_2$ by less than 10\% \cite{roginskii2024nonlinear}. 
Although the absolute values of $\langle\chi^{(3)}\rangle$ are inherently dependent on the experimental setup, our methodology allows one to extract trends and analyze of the evolution of $\chi^{(3)}$ in TeO$_2$-modified glasses relative to SiO$_2$ with respect to the modifier concentration, thereby enabling a deeper understanding of how structural modifications influence NLO properties, independently of the specific experimental configuration. Data for SiO$_2$ models can be found in Ref. (\citenum{roginskii2024nonlinear}).

\begin{table*}[!ht]
    \centering
    \small
    \caption{Refractive index, first and third order non-linear susceptibility calculated for pure TeO$_2$, binary $\rm {(TlO_{0.5})}_{\textit{y}}-{(TeO_2)}_{1-\textit{y}}$ and SiO$_2$ glassy systems. We also provide the diagonal $\chi^3_{d}$ (10$^{-22}$ m$^2$/V$^2$)  and non-diagonal $\chi^3_{nd}$ (10$^{-22}$ m$^2$/V$^2$)  averages of the $\chi^3$ (10$^{-22}$ m$^2$/V$^2$)  matrix, in to addition  $\langle\chi^{(3)}\rangle$. Available experimental data are given between square brackets.}
    \label{tab6_properties_SiO2_Tl2O-TeO2}
    \begin{tabular}{|c||c|c|c|c||c|c|c|c|c|}
        \hline \hline
        Property                     & \multicolumn{4}{c||}{SiO$_2$} & \multicolumn{5}{c|}{$\rm{(TlO_{0.5})}_{\textit{y}}-{(TeO_2)}_{1-\textit{y}}$}                                                                                                                                                                  \\
        \hline
                                     &                               &                                                                                &      &                                       & $y$ = 10\%  & $y$ = 20\%                             & $y$ = 30\%  & $y$ = 40\%                   & $y$ = 50\%  \\
        \hline
        Number                       & 270                           & 408                                                                            & 432  & 456                                   & 456         & 432                                    & 408         & 432                          & 40          \\
        of atoms                     &                               &                                                                                &      &                                       &             &                                        &             &                              &             \\
        \hline \hline
        \hline
        n$_0$                        & 1.49                          & 1.50                                                                           & 1.50 & 1.50 [1.458] \cite{kim1993linear}     & 2.29        & 2.24                                   & 2.21        & 2.17                         & 2.14        \\
        \hline
        $\chi^{(3)}_{d}$             & 1.02                          & 1.19                                                                           & 1.23 & 1.22                                  & 54.38       & 54.26                                  & 55.68       & 52.45                        & 55.52       \\
        $\chi^{(3)}_{nd}$            & 0.41                          & 0.46                                                                           & 0.47 & 0.48                                  & 23.59       & 23.51                                  & 24.38       & 22.44                        & 24.14       \\
        $\chi^{(3)}_{d}/\chi^3_{nd}$ & 2.45                          & 2.59                                                                           & 2.62 & 2.54                                  & 2.30        & 2.31                                   & 2.28        & 2.34                         & 2.3         \\
        $\langle\chi^{(3)}\rangle$   & 1.1                           & 1.27                                                                           & 1.30 & 1.30 [2.095]  \cite{duclere2018third} & 60.90       & 60.78  [48.2, y=25\%]  \cite{dutreilh2003new} & 62.81       & 58.36 [47.6] \cite{dutreilh2003new} & 62.22       \\
        \hline \hline
    \end{tabular}
\end{table*}

\begin{figure}[!htbp]
    \centering
    \includegraphics[width=0.99\linewidth]{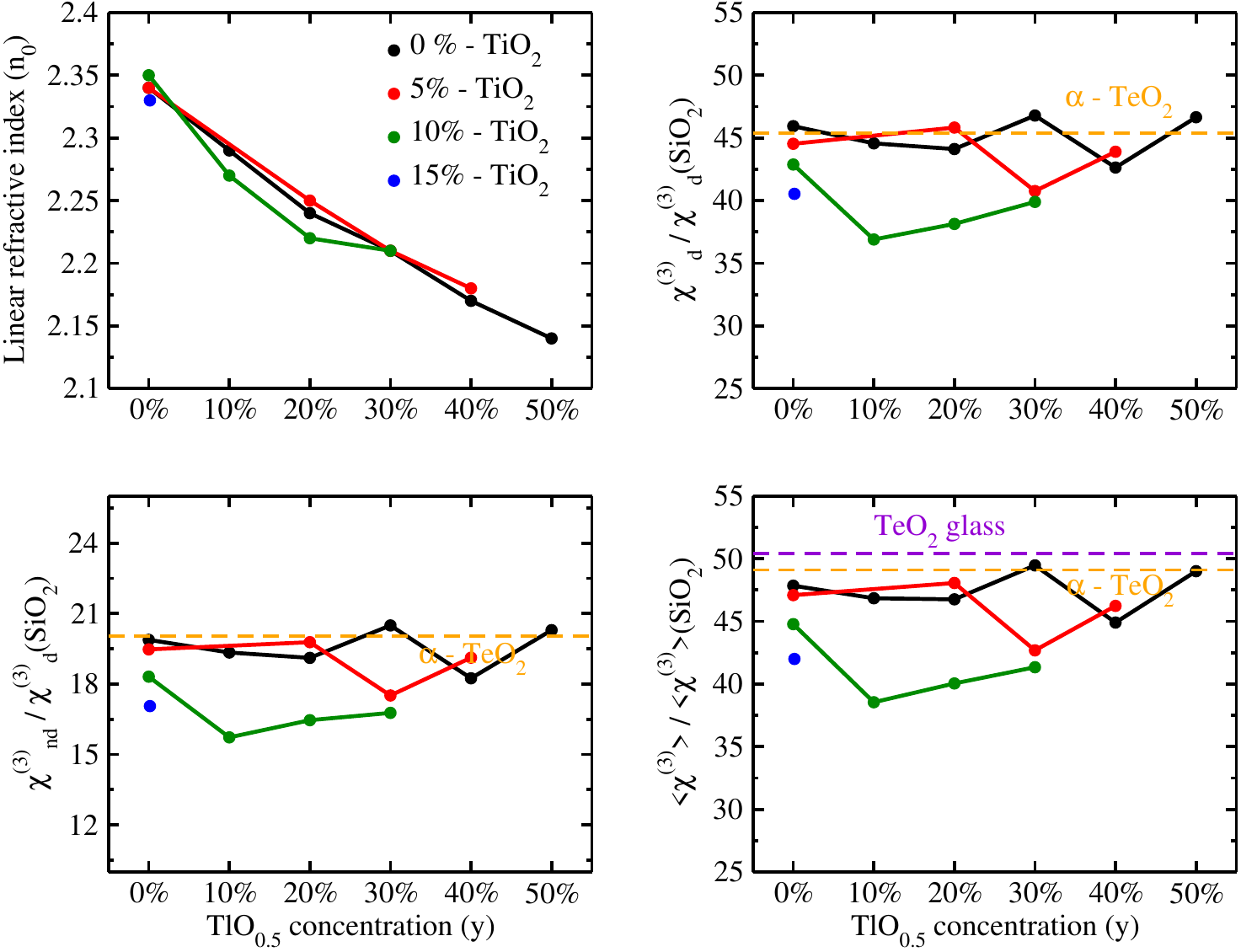}
    \caption{Description of evolution of various linear and non-linear properties in $\rm {(TlO_{0.5})}_{\textit{y}}-{(TeO_2)}_{1-\textit{y}}$, $\rm {(TiO_{2})}_{\text{x}}-{(TeO_2)}_{\text{1-x}}$ and $\rm  {(TiO_{2})}_{\textit{x}}-{(TlO_{0.5})}_{\textit{y}}-{(TeO_2)}_{1-\textit{x}-\textit{y}} $ glasses. Orange color horizontal dashed lines displays the value obtained experimentally for paratellurite \cite{duclere2018third}. Violet color dashed line shows the ratio of $\chi^{(3)}$ in TeO$_2$ to SiO$_2$ glass as observed by Kim et al. \cite{kim1993linear}.}
    \label{label_figure17}
\end{figure}

Refractive indices, third-order non-linear susceptibilities ($\chi^{(3)}_d$, $\chi^{(3)}_{nd}$, and $\langle\chi^{(3)}\rangle$) are reported in Table~\ref{tab6_properties_SiO2_Tl2O-TeO2} for binary $\mathrm{(TlO_{0.5})}_{y}-\mathrm{(TeO_2)}_{1-y}$ glasses and Table~\ref{tab6_titanium_binary_ternary} for ternary $\mathrm{(TiO_2)}_x-\mathrm{(TlO_{0.5})}_y-\mathrm{(TeO_2)}_{1-x-y}$ glasses compared to available experimental data. Figure \ref{label_figure17} shows the various $\chi^{(3)}$ ratios evolution as a function of the composition. For completeness, we also report the results of NLO properties analysis on $\mathrm{(TiO_2)}_x-\mathrm{(TeO_2)}_{1-x}$ binary glasses in Table~\ref{tab6_titanium_binary_ternary} and Figure \ref{label_figure17}, and provide details of the amorphous $\mathrm{(TiO_2)}_x-\mathrm{(TeO_2)}_{1-x}$ glass model generation in the supplementary materials.

Despite a slight overestimation, our calculations yield values of refractive indices in good agreement with experiments (see Tables \ref{tab6_properties_SiO2_Tl2O-TeO2} and \ref{tab6_titanium_binary_ternary}). 
In addition, the refractive index ($n_0$) decreases systematically with the substitution of TeO$_2$ units by modifying agents, such as TiO$_2$ or TlO$_{0.5}$. This reduction is a direct consequence of the reduced polarizability associated with the incorporation of modifier oxides.

Focusing on binary $\rm {(TlO_{0.5})}_{\textit{y}}-{(TeO_2)}_{1-\textit{y}}$, $\langle\chi^{(3)}\rangle$ values remain relatively constant across compositions, in agreement with experimental findings \cite{dutreilh2003new}. Specifically, the ratio of $\chi^{(3)}_d$ and $\chi^{(3)}_{nd}$ in the studied modified tellurite glasses to $\chi^{(3)}_d$ of silicate glasses remain consistently around 45.4 and 20.04, respectively (see Tables~\ref{tab6_properties_SiO2_Tl2O-TeO2} and \ref{tab6_titanium_binary_ternary}). These values are comparable to those reported for paratellurite TeO$_2$ crystal \cite{duclere2018third}. Similarly, the ratio of $\langle\chi^{(3)}\rangle$(A)/$\langle\chi^{(3)}\rangle$(SiO$_2$), where A denotes the modified tellurite glass, also aligns closely with the values observed for paratellurite TeO$_2$ crystals (see Figure \ref{label_figure17}).
These results suggest that the inclusion of TlO$_{0.5}$ modifier into TeO$_2$ host glass has a limited effect on the overall magnitude of $\langle\chi^{(3)}\rangle$, despite promoting the network depolymerization, as discussed in the previous sections. As such, the change in local structure and bonding environment have a minimal effect on the macroscopic NLO properties. This can be related to both, the high stereochemical activity of the lone pair of thallium atoms which induced strong polarisability of thallium oxygen groups \cite{dutreilh2003new}, and the maintain of high enough fraction of BO and the corresponding -Te-O-Te- hyperpolarizable chains.

\begin{table*}[!ht]
    \scriptsize
    \centering
    \caption{Refractive index, first and third order non-linear susceptibility calculated for binary $\rm  {(TiO_{2})}_{\textit{x}}-{(TeO_2)}_{1-\textit{x}}$ and ternary $\rm  {(TiO_{2})}_{\textit{x}}-{(TlO_{0.5})}_{\textit{y}}-{(TeO_2)}_{1-\textit{x}-\textit{y}}$ glassy systems. We also provide the diagonal $\chi^3_{d}$ (10$^{-22}$ m$^2$/V$^2$) and non-diagonal $\chi^3_{nd}$ (10$^{-22}$ m$^2$/V$^2$) averages of the $\chi^3$ (10$^{-22}$ m$^2$/V$^2$)  matrix, in addition to $\langle\chi^{(3)}\rangle$. Available experimental data are given between square brackets.}
    \label{tab6_titanium_binary_ternary}
    \begin{small}
    \begin{tabular}{|c||c|c|c||c|c|c|c|c|c|}
        \hline \hline
        Property                     & \multicolumn{3}{c||}{ $\rm  {(TiO_{2})}_{\textit{x}}-{(TeO_2)}_{1-\textit{x}}$} & \multicolumn{6}{c|}{ $\rm  {(TiO_{2})}_{\textit{x}}-{(TlO_{0.5})}_{\textit{y}}-{(TeO_2)}_{1-\textit{x}-\textit{y}}$}                                                                                                                                                                                                     \\
        \hline
                & $x$ = 5\%    & $x$ = 10\%      & $x$ = 15\%     & $x$ = 5\% & $x$ = 5\%   & $x$ = 5\%  & $x$ = 10\% & $x$ = 10\%  & $x$ = 10\%  \\
                &              &                 &                & $y$ = 20\% &  $y$ = 30\%   & $y$ = 40\% & $y$ = 10\%  & $y$ = 20\%   &  $y$ = 30\% \\
        \hline
        Number                       & 480                                                                         & 480                                                                                                                & 480                      & 540                 & 510                                & 480                  & 456                    & 540                   & 510                   \\
        of atoms                     &                                                                                  &                                                                                                                      &                               &                       &                                        &                       &                        &                        &                        \\
        \hline \hline
        \hline
        n$_0$                        & 2.34                                                                             & 2.35                                                                                                                 & 2.33                          & 2.25                  & 2.21 [2.12] \cite{udovic2009formation} & 2.18                  & 2.27                   & 2.22                   & 2.21                   \\
        \hline
        $\chi^{(3)}_{d}$             & 54.33                                                                            & 52.32                                                                                                                & 49.46                         & 55.91                 & 49.73                                  & 53.56                 & 45.02                  & 46.53                  & 48.68                  \\
        $\chi^{(3)}_{nd}$            & 23.75                                                                            & 22.34                                                                                                                & 20.81                         & 24.13                 & 21.36                                  & 23.33                 & 19.19                  & 20.08                  & 20.46                  \\
        $\chi^{(3)}_{d}/\chi^3_{nd}$ & 2.29                                                                             & 2.34                                                                                                                 & 2.38                          & 2.32                  & 2.33                                   & 2.30                  & 2.35                   & 2.32                   & 2.38                   \\
        $\langle \chi^{(3)} \rangle$ & 61.22                                                                            & 58.20  (42.2) \cite{dutreilh2003new}                                                                                        & 54.61  [39.6] \cite{dutreilh2003new} & 62.48                 & 55.49                                  & 60.10                 & 50.16                  & 52.05                  & 53.74                  \\
        \hline \hline
    \end{tabular}
\end{small}
\end{table*}

Coming to $\mathrm{(TiO_2)}_x-\mathrm{(TeO_2)}_{1-x}$ binary glasses, the calculated $\langle\chi^{(3)}\rangle$ values exhibit a monotonous decrease with increasing TiO$_2$ content, consistent with experimental observations \cite{dutreilh2003new}. This trend can be attributed to the substitution of TeO$_2$ units with TiO$_2$, which reduces the material's hyperpolarizability. Unlike Te atoms, Ti atoms lack electronic lone pairs, leading to a diminished cumulative contribution to $\chi^{(3)}$ from lone-pair-induced hyperpolarizabilities in the glass matrix. In addition, the insertion of Ti atoms in the -Te-O-Te- chains leads to the formation of -Te-O-Ti- bridges which may decrease the overall hyperpolarizability of these chains.

The trends observed in $\mathrm{(TiO_2)}_x-\mathrm{(TeO_2)}_{1-x}$ binary glasses also extend to the  systems where the interplay between TiO$_2$ and TlO$_{0.5}$ modifiers further influences the structural and optical properties. 

Finally, looking at the ternary $\mathrm{(TiO_2)}_x-\mathrm{(TlO_{0.5})}_y-\mathrm{(TeO_2)}_{1-x-y}$ glass, Figure~\ref{label_figure17} suggests that high concentrations of TiO$_2$ reduces the nonlinear optical properties. However, by incorporating appropriate fractions of TlO$_{0.5}$ and TiO$_2$ modifiers to the parent TeO$_2$ glass, the nonlinear optical properties can be maintained close to that of the parent glass while preserving the overall network connectivity of the material \cite{udovic2009formation}. Specifically, we observe that, irrespective of the TlO$_{0.5}$ concentration, introducing $x = $5\% TiO$_2$ yields $\langle\chi^{(3)}\rangle$(A)/$\langle\chi^{(3)}\rangle$(SiO$_2$) ratios of approximately 50, comparable to those obtained in the case of binary $\mathrm{(TlO_{0.5})}_y-\mathrm{(TeO_2)}_{1-y}$ glasses. However, increasing the TiO$_2$ concentration to $x = $10\% reduces this ratio to approximately 40, highlighting the trade-off between TiO$_2$ content and nonlinear optical performance.

\section{Conclusions}

In this work, we used \textit{ab-initio} molecular dynamics to perform a comprehensive comparison of $\rm{(TlO_{0.5})}_{\textit{y}}-{(TeO_2)}_{1-\textit{y}}$ binary and $\rm  {(TiO_{2})}_{\textit{x}}-{(TlO_{0.5})}_{\textit{y}}-{(TeO_2)}_{1-\textit{x}-\textit{y}} $ ternary tellurite glasses.  For both systems, our models reproduce fairly the available experimental X-ray PDF, thereby validating the achieved structural models.

In binary glasses, as the thallium modifier content $y$ increases, we observe a progressive depolymerization of the Te–O network: the average Te coordination number $n_{\rm Te}$ falls, bridging Te–O–Te linkages are replaced by Te=O$^{-}\cdots$Tl$^{+}$ units, and the non-bridging oxygen concentration rises. Ring‐analysis further reveals that Tl$^{+}$ does not participate directly in ring formation. Instead, the growing modifier fraction opens up smaller Te–O rings, shifting the ring‐size distribution towards larger membered rings and confirming the loss of network connectivity. 
In ternary glasses, Ti$^{4+}$ acts as a network former: increasing $x$ at fixed thallium content raises the Te–O coordination number, enhances the fraction of 4- and 5-fold Te sites at the expense of 3-fold motifs, and promotes the conversion of non-bridging to bridging oxygens. Rings analysis confirms that Ti‐containing rings are smaller and more numerous, indicating a re-polymerization of the network and suggesting improved mechanical rigidity compared to the binary glasses.

A comparison between the computed Raman spectra on our large fully periodic systems, revealed an outstanding agreement with experimental data where the structural depolymerization was tracked by following the Raman bands shift as a function of the glass composition. As for the non-linear optical properties, our calculations indicate that the $\langle\chi^{(3)}\rangle$ values remain nearly constant as the concentration of TlO$_{0.5}$ increases in binary glasses in agreement with experimental observations. Additionally, when referenced to $\chi^{(3)}_d$ computed on pure SiO$_2$ systems, the ratio of $\chi^{(3)}_d$ and $\chi^{(3)}_{nd}$ in the studied modified tellurite glasses are found around 45.4 and 20.04, respectively, which corresponds to values previously observed in paratellurite TeO$_2$ crystals \cite{duclere2018third}. The addition of a small amount (5 \%) of TiO$_2$ preserves the high nonlinearity of the pure TeO$_2$ network, while further maintaining the glass refractive index and $\chi^{(3)}$ to values in excellent agreement with experiment. 

Overall, our study quantitatively clarifies how TlO$_{0.5}$ and TiO$_2$ modifiers tailor the atomic structure and nonlinear properties of tellurite glasses, and provides a predictive framework for optimizing TeO$_2$-based materials for photo-optical applications. By balancing depolymerization (via Tl$^{+}$) and re-polymerization (via Ti$^{4+}$), one can finely tune both the network connectivity and the NLO response in these glassy systems.

\section*{Acknowledgments}
This work was supported by the ANR via the TRAFIC project (ANR-18-CE08-0016-01) and by r\'egion Nouvelle Aquitaine via the F2MH project AAP NA 2019-1R1M01. Calculations were performed by using resources from Grand Equipement National de Calcul Intensif (GENCI, grants No. AX0910832 and AX0913426). We are grateful to TGCC for the generous CPU allocation within the special session project spe00019. We used computational resources provided by the computing facilities M\'esocentre de Calcul Intensif Aquitain (MCIA) of the Universit\'e de Bordeaux and of the Universit\'e de Pau et des Pays de l'Adour.

\bibliographystyle{apsrev4-1}
\bibliography{biblio.bib}

\end{document}